\documentstyle[aps, floats, epsfig]{revtex}
\begin{document}
\draft
 
\hsize\textwidth\columnwidth\hsize\csname@twocolumnfalse\endcsname
\title{Dynamic Cosmic Strings I}

\author{K R P Sj\"odin\thanks{E-mail: K.R.Sjodin@maths.soton.ac.uk}, 
U Sperhake\thanks{E-mail: U.Sperhake@maths.soton.ac.uk} and 
J A Vickers\thanks{E-mail: J.A.Vickers@maths.soton.ac.uk}}
\address{
Faculty of Mathematical Studies, \\
University of Southampton, \\ 
Southampton, S017 1BJ, U.K.
}
\date{10 February 2000}
\maketitle

\begin{abstract}

The field equations for a time dependent cylindrical cosmic string
coupled to gravity are reformulated in terms of geometrical variables
defined on a $2+1$-dimensional spacetime by using the method of Geroch
decomposition. Unlike the 4-dimensional spacetime the reduced case is
asymptotically flat. A numerical method for solving the field
equations which involves conformally compactifying the space and
including null infinity as part of the grid is described. It is shown
that the code reproduces the results of a number of  vacuum solutions
with one or two degrees of freedom. In the final section the interaction
between the cosmic string and a pulse of gravitational radiation is
briefly described. This is fully analyzed in the sequel.
\end{abstract}
\pacs{PACS number(s): 0420.Ha, 0420.Jb, 0425.Dm, 04.30Db}

\section{Introduction}

Cosmic strings are topological defects that formed during phase
transitions in the early universe. They are important because they
are predicted by grand unified theories and produce  density
perturbations in the early universe that might be important
in the formation of galaxies and other large scale structures \cite{[1]}. They
are also important since they are thought to be sources of
gravitational radiation due to rapid oscillatory motion \cite{[2]}. In the
simplest case of a string moving in a fixed background one can take
the thin string limit and the dynamics are given by the
Nambu--Goto action \cite{[3]} which is known to admit oscillatory
solutions. However in order to fully understand the behavior of
cosmic strings one should study the field equations for a cosmic
string coupled to Einstein's equations.

A cosmic string is described by a $\mathrm{U(1)}$ gauge vector field
$A_\mu$ coupled to a complex scalar field $\Phi=1/\sqrt 2
Se^{i\phi}$. The Lagrangian for these coupled fields is given by
\begin{equation}
  L_M=\frac{1}{2}\nabla_\mu S\nabla^\mu S + \frac{1}{2}S^2(\nabla_\mu\phi+eA_\mu)(\nabla^\mu\phi+eA^\mu) 
      - \lambda(S^2-\eta^2)^2 - \frac{1}{4}F_{\mu\nu}F^{\mu\nu},\label{1.1}
\end{equation}  
where $F_{\mu\nu}=\nabla_\mu A_\nu-\nabla_\nu A_\mu$, $e$,
$\lambda$ are positive coupling constants and $\eta$ is the vacuum 
expectation value. 

The Einstein-scalar-gauge field equations for an infinitely long
static cosmic string have been investigated by Laguna and Garfinkle  
\cite{lagunaI}. 
Some geometrical techniques related to those used in this paper have also been
used in the analysis of cosmic strings by Peter and Carter
\cite{Peter} and by Carter \cite{Carter}.
In the present paper (and its sequel) we will
investigate the behavior of a time dependent cylindrical cosmic
string coupled to gravity. In particular we investigate the effect of
a pulse of gravitational radiation on an initially static cosmic
string and the corresponding gravitational radiation that is emitted
as a result of oscillations in the string. Since we are interested in
the gravitational radiation produced by the string it is desirable to
measure this at null infinity where the gravitational flux is
unambiguously defined and one does not need to impose artificial
outgoing radiation conditions at the edge of the numerical grid. However the
infinite length of the string  in the $z$-direction prevents the
spacetime from being asymptotically flat. We therefore follow the
approach of Clarke et al. \cite{CDV} and use  a Geroch decomposition
with respect to the Killing vector in the $z$-direction to reformulate
the problem in $2+1$ dimensions. Note however that unlike Clarke et al.
\cite{CDV} we apply the Geroch decomposition to the entire problem not
just to the exterior characteristic region. The gravitational degrees 
of freedom of the $3+1$ problem are then encoded in two geometrically defined
scalar fields defined on the $2+1$-dimensional spacetime. These are the norm
of the Killing vector $\nu$ and the Geroch potential $\tau$ for the rotation.
We show in section III that the energy-momentum tensor of these
fields describes the gravitational energy of the original cylindrical
problem. An important feature of the reduced $2+1$ spacetime is that
it is asymptotically flat and this allows us to conformally compactify
the spacetime and include null infinity as part of the numerical grid.   

In section II we briefly describe the Geroch decomposition and the
field equations that one obtains in the cylindrical case. We also show
how it is possible to rescale $t$, $\rho$ and the matter variables to
simplify the equations. In order to demonstrate the numerical accuracy
of the full code it is useful to compare it with either an exact solution   
or else with some other independently produced numerical
results. Unfortunately this is not possible for the full code where we
must satisfy ourselves with the internal consistency of the code as
demonstrated by convergence testing for example. If one
considers only the gravitational part of the code it is possible to
specialize to the case of a vacuum solution. The simplest of these is the
Weber--Wheeler solution \cite{weber}, but we also consider a rotating
vacuum solution due to Xanthopoulos \cite{Xan} which describes a
thin cosmic string and a second rotating vacuum solution due to Piran
et al. \cite{Piran}. In order to compare our results with
these exact solutions we must first write them in terms of the
Geroch potential, this is done in section IV. This description is
useful not just for numerical purposes but also
in interpreting these solutions since the two polarization states for
the gravitational radiation have a simple interpretation in terms of
the Geroch variables $\nu$ and $\tau$.  The numerical code used in
this case, the convergence analysis and the comparison between the
numerical results and the exact solution is given in section IV. As
far as the matter part of the code is concerned there are no exact
solutions and one must compare the code with other numerical results. 
The simplest special case involves  decoupling the matter variables 
from the metric variables and considering the equations of motion in
Minkowski space. There do not seem to be other results available in the 
dynamic case, but the static solutions have been investigated by a
number of authors \cite{sstringI}, \cite{sstringII} and \cite{sstringIII}. 
Another solution which has been
investigated previously is a static string which is coupled to the
gravitational field. Finding such solutions is much harder than one
might suppose due to the asymptotic behavior of the matter
variables. As well as the physical solution to the equations there is
an exponentially diverging non-physical solution \cite{sstringIII} which
must be suppressed. By compactifying the radial coordinate we can
control the behavior of the solution at infinity in the static case
by using a relaxation scheme. This allows us to obtain
solutions for all values of the radius rather than the fairly
restricted range of $\rho$ that had been previously obtained, and also
permits us to use the proper boundary conditions at infinity. The static 
string is described in further detail in sections V and VI. Finally we
briefly describe the numerics for the dynamic string coupled
to gravity. Here again the asymptotics of the matter variables make it
hard to write a stable code, but by using an implicit scheme we are
able to produce a code with long term stability and second order
convergence which agrees with the previous results in the special
cases described earlier. From a physical point of view the most
interesting feature of this code is that it is able to describe the
interaction of a gravitational field with two degrees of freedom with
the full non-linear cosmic string equations. So far we have only investigated
the interaction of the cosmic string with an incoming
Weber--Wheeler type pulse of gravitational radiation with just one
degree of freedom. We
find that the pulse excites the cosmic string and causes the scalar
and vector fields  to vibrate
with a frequency which is roughly proportional to their respective
masses. This oscillation slowly decays and the string eventually 
returns to its previous static state. This is briefly described in section VII 
and in detail in the sequel \cite{paperII}, 
which we henceforth will refer to as paper II, where  comparisons with
other results and a full convergence analysis is given.

\section{The Geroch decomposition}

As we explained above it is not possible for a cylindrically symmetric
cosmic string to be asymptotically flat due to the infinite extent of
the string in the $z$-direction. By factoring out this
direction we can obtain a 3-dimensional spacetime that is
asymptotically flat. If the Killing vector in the $z$-direction is
hypersurface orthogonal then one can simply project onto the surfaces
${\cal S}$ given by $z=\mathrm{const}$. 
However we wish to consider cylindrical solutions which
also have a rotating mode and in this case the Killing vector
$\xi^\mu$ is not hypersurface orthogonal. Geroch \cite{geroch}
has shown how to factor out the Killing direction in this more general
case. The idea is to identify points which lie on the same integral
curves of the Killing vector field and thus obtain ${\cal S}$ as a
quotient space rather than as a subspace. There is then a
one-to-one correspondence between tensor fields on ${\cal S}$ and
tensor fields on the 4-dimensional manifold $M$ which have vanishing 
contraction with the Killing vector and also vanishing Lie derivative
along $\xi^\mu$.
One may therefore use the four dimensional metric $g_{\mu\nu}$ to
define a metric $h_{\mu\nu}$ on ${\cal S}$ according to the equation  
\begin{equation}
  h_{\mu\nu}=g_{\mu\nu}+(\xi^{\sigma}\xi_{\sigma})^{-1}\xi_\mu\xi_\nu.
\end{equation}
The extra information in the 4-metric is encoded in two new geometric 
variables; the norm of the Killing vector
\begin{equation}
  \nu=-\xi^\mu\xi_\mu
\end{equation}
(where we have introduced the minus sign to make $\nu$ positive in the
spacelike case) and the twist
\begin{equation}
  \tau_\mu=-\epsilon_{\mu\nu\tau\sigma}\xi^\nu\nabla^\tau\xi^\sigma.
\end{equation}
Geroch then showed how it is possible to rewrite Einstein's equations
in terms of the 3-dimensional Ricci curvature of $({\cal S}, h)$ and
equations involving the 3-dimensional covariant derivatives of
$\nu$ and the twist.
If we let $D_\mu$ define the covariant derivative with respect to
$h_{\mu\nu}$ then one can show that
\begin{equation}
  D_{[\rho}\tau_{\sigma]}=\epsilon_{\rho\sigma\mu\nu}\xi^\mu 
R^\nu_\tau\xi^\tau,
\end{equation}
where $R^\nu_\tau$ is the 4-dimensional Ricci tensor. 
It is clear that this vanishes in vacuum so that $\tau_\sigma$ is curl
free and may be defined in terms of a potential. It is a
remarkable fact that this remains true for spacetimes with a cosmic
string energy-momentum tensor so that even in this case we may write
\begin{equation}
  \tau_a=D_a\tau,
\end{equation}
where we have introduced the convention of using Latin indices to
describe quantities defined on ${\cal S}$.

We may now write Einstein's equations for the 4-dimensional spacetime $(M,g)$
in terms of the Ricci curvature of $({\cal S},h)$ and the two scalar
fields $\nu$ and $\tau$ defined on ${\cal S}$. We obtain
\begin{eqnarray}
    {\cal R}_{ab} &=&
\frac{1}{2}\nu^{-2}[(D_a\tau)(D_b\tau)-h_{ab}(D_m\tau)
(D^m\tau)] + \frac{1}{2}\nu^{-1}D_aD_b\nu - 
\frac{1}{4}\nu^{-2}(D_a\nu)(D_b\nu)\nonumber\\
 &&+8\pi h_a{}^\mu h_b{}^\nu
(T_{\mu\nu}-\frac{1}{2}g_{\mu\nu}T),\label{D90}\\
D^2 \nu &=& \frac{1}{2}\nu^{-1}(D_m\nu)(D^m\nu) -
\nu^{-1}(D_m\tau)(D^m\tau)
+16\pi (T_{\mu\nu}-\frac{1}{2}g_{\mu\nu}T)\xi^\mu\xi^\nu,\label{D91}\\
D^2 \tau &=& \frac{3}{2}\nu^{-1}(D_m\tau)(D^m\nu)\label{D92}.
\end{eqnarray}

The transformation to the 3-dimensional description is not only 
mathematically convenient but is physically meaningful.
If the Killing vector is hypersurface orthogonal then $\tau$ vanishes
and the gravitational radiation has only one polarisation which may be
defined in a simple way in terms of $\nu$. If there are both
polarisations present then the $+$ mode is given in terms of
$\nu$ while the $\times$ mode is given in terms of $\tau$ [see
equation (\ref{polar}) below]. One can simplify things by making a conformal
transformation and using the metric ${\tilde h}_{ab}=\nu h_{ab}$ in
which case we can write the vacuum Einstein--Hilbert Lagrangian on $M$ in
terms of 3-dimensional variables on ${\cal S}$
\begin{eqnarray}
I_G&=&\int_M R \sqrt{-g} d^4x \\ 
   &=& \int_{\cal S}\{\tilde{\cal R} -\frac{1}{2}\nu^{-2}
[{\tilde h}^{ab}(\tilde
   D_a\tau)(\tilde D_b\tau)+{\tilde h}^{ab}(\tilde
   D_a\nu)(\tilde D_b\nu)]\}\sqrt{\tilde h}d^3x
\end{eqnarray}
and we see that the 4-dimensional gravitational field is described in
three dimensions by the two scalar fields $\nu$ and $\tau$ conformally
coupled to the 3-dimensional spacetime with metric ${\tilde
h}_{ab}$. Since the 3-dimensional spacetime has no Weyl curvature it 
is essentially non-dynamic and we see
that $\nu$ and $\tau$ encode the two gravitational degrees of freedom
in the original spacetime. The corresponding `energy-momentum' tensor
for these fields is
\begin{equation}
{\hat T}_{ab}=\frac{1}{2}\nu^{-2}[{\tilde D}_a\tau{\tilde
D_b}\tau-\frac{1}{2}{\tilde h}_{ab}{\tilde h}^{cd}(\tilde
   D_c\tau)(\tilde D_d\tau)+{\tilde D}_a\nu{\tilde
D_b}\nu-\frac{1}{2}{\tilde h}_{ab}{\tilde h}^{cd}(\tilde
   D_c\nu)(\tilde D_d\nu)]
\end{equation}
and if there is matter present in four dimensions there are also
additional matter terms in three dimensions. As shown in equation
(\ref{energy}) this 3-dimensional `energy-momentum' tensor for $\nu$ and
$\tau$ gives the correct expression for the 4-dimensional
gravitational energy.

\section{The field equations}

For the case of a cylindrically symmetric vacuum spacetime one can
write the metric in Jordan, Ehlers, Kundt and Kompaneets (JEKK) form
\cite{JEK}, \cite{Komp}
\begin{equation} 
  ds^2 = e^{2(\gamma-\psi)}(dt^2 - d\rho^2) - \rho^2e^{-2\psi}d\phi^2 
         - e^{2\psi}(\omega d\phi+dz)^2.
\end{equation}
However this form of the metric is not compatible with the cosmic
string energy momentum tensor so we follow Marder \cite{marder} by
introducing an extra variable $\mu$ into the metric and writing 
it in the form
\begin{equation}
  ds^2 = e^{2(\gamma-\psi)}(dt^2 - d\rho^2) - \rho^2e^{-2\psi}d\phi^2 
        - e^{2(\psi+\mu)}(\omega d\phi+dz)^2. \label{rmetric}
\end{equation}
This form of the line element enables us to make easy comparisons with 
the JEKK vacuum solutions previously considered numerically by Dubal
et al. \cite{CDD} and d'Inverno et al. \cite{DDS}. 
The field equation for $\mu$ decouples from those for the other metric
variables and it has a source term given by $T_{00}-T_{11}$. The
physical interpretation of $\mu$ is briefly discussed by Marder \cite{marder}.
In the static case one can show that $C^2$ regularity on the axis
implies that $\mu=\ln(\nu)-\gamma$. The metric given by (\ref{rmetric})
has zero shift and lapse
determined by the condition $g_{tt}=-g_{\rho\rho}$. In this gauge the null
geodesics are given by the simple conditions $u=t-\rho=\mathrm{const.}$ and
$v=t+\rho=\mathrm{const}$. The remaining coordinate freedom is given by the
freedom to relabel the radial null surfaces: $u \to f(u)$
and $v \to g(v)$ where $f$ and $g$ are arbitrary functions. We may fix
this by specifying the initial values of $\mu$ and its derivative.  For example
we can choose $\mu$ to be equal to its static value and $\mu_{,t}$ to
vanish, but due to time dependent matter source terms in the evolution 
equation for $\mu$ [see equation (\ref{E3}) below] this does not 
make $\mu$ constant in time.

In terms of these variables we find the norm of the Killing vector in
the $z$-direction $\xi^\mu=\delta^\mu_3$ to be given by
\begin{equation}
  \nu=e^{2(\psi+\mu)}
\end{equation}
and the twist potential is related to $\omega$ by
\begin{equation}
  D_{\sigma}\tau=\rho^{-1}e^{4\psi+3\mu}(\omega_{,\rho},\omega_{,t}, 0,0).
\end{equation}
Finally the conformal 3-metric ${\tilde h}_{ab}$ is given by
\begin{equation}
  d{\tilde\sigma}^2=e^{2(\gamma+\mu)}(dt^2-d\rho^2)-\rho^2e^{2\mu}d\phi^2.
\end{equation}

It is also of interest to calculate ${\hat T}_{ab}$ in terms of these
variables. We find
\begin{equation}
  {\hat T}_{ab}t^at^b=\frac{1}{8}e^{-2(\gamma+\mu)}\left[\left(\frac{\nu_{,u}}{\nu}\right)^2
  +\left(\frac{\nu_{,v}}{\nu}\right)^2+\left(\frac{\tau_{,u}}{\nu}\right)^2
  +\left(\frac{\tau_{,v}}{\nu}\right)^2\right],
\end{equation}
where $t^a$ is a unit timelike vector proportional to
$\frac{\partial}{\partial t}$. Note that the quantities
\begin{equation}
  A=\left(\frac{\nu_{,u}}{\nu}\right)^2, \qquad
  B=\left(\frac{\nu_{,v}}{\nu}\right)^2, \qquad
  C=\left(\frac{\tau_{,u}}{\nu}\right)^2, \qquad
  D=\left(\frac{\tau_{,v}}{\nu}\right)^2,
 \label{polar}
\end{equation} 
are exactly the same as the quantities $A$, $B$, $C$ and $D$ which are
given (by more complicated expressions) in terms of $\psi$ and
$\omega$ in equations (4a)--(4d) of Piran et al. \cite{PiranII} and describe
the two polarizations of the cylindrical gravitational
field. Furthermore if we consider the special case of vacuum solutions
and integrate ${\hat T}_{ab}t^at^b$ over the region 
\hbox{$V=\{0\leq \rho \leq \rho_0,\, t=t_0\}$} 
with respect to the volume form $dV$ on $t=t_0$ we find
\begin{eqnarray}
E(t_0, \rho_0)&=&\int\int_V {\hat T}_{ab}t^at^bdV \\
           &=&\frac{\pi}{4}\int_0^{\rho_0}
e^{-\gamma}\left[\left(\frac{\nu_{,u}}{\nu}\right)^2
+\left(\frac{\nu_{,v}}{\nu}\right)^2+\left(\frac{\tau_{,u}}{\nu}\right)^2
+\left(\frac{\tau_{,v}}{\nu}\right)^2\right]\rho d\rho \\
&=&2\pi\int_0^{\rho_0}\gamma_{,\rho}e^{-\gamma}d\rho \\
&=& 2\pi[1-e^{-\gamma(t_0,\rho_0)}],
\label{energy}
\end{eqnarray}
where we have used the vacuum field equations for $\gamma_{,\rho}$ [see
equation (\ref{gam2}) below]. Note that this is the same as the energy
obtained by Ashtekar et al. \cite{ashtekar} but does not require the
Killing vector to be hypersurface orthogonal. It differs from the
C-energy in general but agrees with it in the linearized case.
   
As far as the matter variables are concerned we make the obvious
generalization of the form used by Garfinkle \cite{garfinkle} and
write
\begin{eqnarray}
  \Phi&=&\frac{1}{\sqrt 2}S(t,\rho)e^{i\phi}, \\
  A_\mu&=&\frac{1}{e}[P(t,\rho)-1]\nabla_\mu\phi.
\end{eqnarray}
We may now write the field equations for the complete system. Since we
are working in three dimensions we have three independent evolution
equations. After some algebra these may be written as
\begin{eqnarray}
&&\Box\nu - \nu^{-1}(\tau_{,t}^2-\tau_{,\rho}^2-\nu_{,t}^2 
+\nu_{,\rho}^2)-\mu_{,t}\nu_{,t}+\mu_{,\rho}\nu_{,\rho}\nonumber\\
&&=-8\pi\nu[2\lambda \nu^{-1}e^{2(\gamma+\mu)}(S^2-\eta^2)^2 
+ e^{-2}\rho^{-2}\nu e^{-2\mu}(P_{,t}^2-P_{,\rho}^2)],\label{E1}\\
&&\Box\tau + 2\nu^{-1}(\tau_{,t}\nu_{,t}-\tau_{,\rho}\nu_{,\rho}) 
- (\mu_{,t}\tau_{,t}-\mu_{,\rho}\tau_{,\rho})=0,\label{E2}\\
&&\Box\mu+\rho^{-1}\mu_{,\rho}-\mu_{,t}^2+\mu_{,\rho}^2=-8\pi[2\lambda\nu^{-1}e^{2(\gamma+\mu)}(S^2-\eta^2)^2 +
\rho^{-2}e^{2\gamma}S^2P^2],\label{E3}
\end{eqnarray}
where $\Box$ represents the flat spacetime d'Alembertian which in
cylindrical coordinates is given by
\begin{equation}
  \Box=  - \frac{\partial^2}{\partial t^2} + 
  \frac{\partial^2}{\partial\rho^2} + \rho^{-1}\frac{\partial}{\partial\rho}.
  \label{dal}
\end{equation}

There are also three constraint equations, one of which vanishes
identically due to the rotational symmetry. The remaining equations
give
\begin{eqnarray}
  \gamma_{,t} &=& \frac{\rho}{1+\rho \mu_{,\rho}} \left[ \mu_{,t\rho}
              -  \mu_{,t}(\gamma_{,\rho} + \mu_{,\rho})
              +\frac{1}{2}\nu^{-2}(\tau_{,t}\tau_{,\rho} + \nu_{,t} \nu_{,\rho})
              +8\pi (S_{,t} S_{,\rho} + e^{-2} \rho^{-2}\nu e^{-2\mu}
              P_{,t} P_{,\rho})\right],
              \\
  \gamma_{,\rho} &=& \,\frac{\rho}{\rho^2\mu_{,t}^2-(1+\rho\mu_{,\rho})^2} \left(
              (1+\rho\mu_{,\rho}) \left\{ \vphantom{\frac{\rho}{2\nu^2}}
              -4\pi [
              2\nu^{-1} e^{2(\gamma+\mu)}\lambda (S^2-\eta^2)^2 +(S_{,t}^2
              +S_{,\rho}^2) \nonumber
	      \right. \right. \\
           && \left.
              + e^{-2}\rho^{-2}\nu e^{-2\mu}  (P_{,t}^2+P_{,\rho}^2)
              + \rho^{-2} e^{2\gamma} S^2 P^2] -\frac{1}{4}\nu^{-2}
              (\tau_{,t}^2+\tau_{,\rho}^2+\nu_{,t}^2+\nu_{,\rho}^2)
              - \rho^{-1}\mu_{,\rho} -\mu_{,\rho\rho} \right\}
              \nonumber \\
           && \left.
              +\rho\mu_{,t} \left[
              \frac{1}{2}\nu^{-2}(\tau_{,t}\tau_{,\rho}+\nu_{,t}\nu_{,\rho})
              + \rho^{-1}\mu_{,t}+\mu_{,t\rho} +8\pi (S_{,t}S_{,\rho}
              + e^{-2}\rho^{-2}\nu e^{-2\mu}P_{,t} P_{,\rho})
              \right] \right).
\label{gam2}
\end{eqnarray}
Finally there are the equations for the matter variables $S$ and $P$
which may be derived from the Euler--Lagrange equations for $L_M$ or
alternatively from the contracted Bianchi identities.
\begin{eqnarray}
&&\Box S - S_{,t}\mu_{,t} + S_{,\rho}\mu_{,\rho} = S[4\lambda 
\nu^{-1}e^{2(\gamma+\mu)}(S^2-\eta^2)+\rho^{-2}e^{2\gamma}P^2],\label{E1aa}\\
&&\Box P - P_{,t}(\nu^{-1}\nu_{,t}-\mu_{,t}) 
+ P_{,\rho}(\nu^{-1}\nu_{,\rho}-\mu_{,\rho}-2\rho^{-1}) 
= e^2 \nu^{-1}e^{2(\gamma+\mu)}PS^2. \label{E1a}
\end{eqnarray}

We also need to supplement these equation by boundary conditions on
the axis. For the 4-dimensional metric variables the simplest
condition is to require the metric
to be $C^2$ on the axis so that we have a well defined curvature
tensor. This gives the conditions
\begin{eqnarray}
  \psi(t,\rho) &=&a_1(t)+O(\rho^2), \\  
  \omega(t,\rho) &=&O(\rho^2), \\
  \mu(t,\rho) &=& a_2(t)+O(\rho^2), \\
  \gamma(t,\rho)&=&O(\rho).
\end{eqnarray}
In terms of $\nu$ and $\tau$ this gives
\begin{eqnarray}
  \nu(t,\rho) &=& a_3(t)+O(\rho^2),\\
  \tau(t,\rho)&=&O(\rho^2),
\end{eqnarray}
where we have chosen the additive constant in the definition of the
potential $\tau$ so that it vanishes on the axis. In certain
situations $C^2$ regularity
is too strong and one must impose the weaker condition of elementary
flatness \cite{elementary}. However even this is too strong for the 
Xanthopoulos solution which has a conical singularity on the axis.  

The boundary conditions for $S$ and $P$ on the axis are \cite{garfinkle}
\begin{eqnarray}
  S(t,\rho) &=& O(\rho),\\   
  P(t,\rho) &=& 1+O(\rho^2).
\end{eqnarray}

In order to consider the behavior of the solution at null infinity we
first transform from $t$ to a null time coordinate $u=t-\rho$. We also wish
to compactify the region and following Clarke et al. \cite{CDV} 
we make the transformation
\begin{equation} 
  y=\frac{1}{\sqrt \rho}.
\end{equation}
In terms of these variables the equations become
\begin{eqnarray}
    &&\Box\nu + y^3\nu^{-1}(\tau_{,u}\tau_{,y}-\nu_{,u}\nu_{,y})
+\frac{1}{4}y^6\nu^{-1}(\tau_{,y}^2-\nu_{,y}^2)
+\frac{1}{2}y^3(\mu_{,y}\nu_{,u}+\mu_{,u}\nu_{,y}
+\frac{1}{2}y^3\mu_{,y}\nu_{,y})\nonumber\\
    &&=8\pi\nu[-2\lambda \nu^{-1}e^{2(\gamma+\mu)}(S^2-\eta^2)^2 
+ \frac{1}{2}e^{-2}y^7\nu e^{-2\mu}(\frac{1}{2}y^3P_{,y}^2
+2P_{,u}P_{,y})],\label{compact1}\\
    &&\Box\tau - y^3\nu^{-1}(\tau_{,u}\nu_{,y}+\tau_{,y}\nu_{,u}
+\frac{1}{2}y^3\tau_{,y}\nu_{,y}) + \frac{1}{2}y^3
(\mu_{,y}\tau_{,u}
+\mu_{,u}\tau_{,y}+\frac{1}{2}y^3\mu_{,y}\tau_{,y})=0,\\
    &&\Box\mu-y^2(\mu_{,u}+\frac{1}{2}y^3\mu_{,y})-
\frac{1}{2}y^3\mu_{,y}(\frac{1}{2}y^3\mu_{,y}-2\mu_{,u})\nonumber\\
    &&=-8\pi[2\lambda\nu^{-1}e^{2(\gamma+\mu)}(S^2-\eta^2)^2 
+ y^4e^{2\gamma}S^2P^2],
\end{eqnarray}
\begin{eqnarray}
\gamma_{,u}&&=y^2\textbf{(}(1-\frac{1}{2}y\mu_{,y})
[-\frac{1}{2}y^3\mu_{,uy}-\mu_{,uu}+\mu_{,u}^2-
\frac{1}{2}\nu^{-2}(\tau_{,u}^2+\frac{1}{2}y^3\tau_{,u}\tau_{,y}
+\nu_{,u}^2+\frac{1}{2}y^3\nu_{,u}\nu_{,y})]\nonumber\\
    &&+\frac{1}{8}y^3\mu_{,u}[\mu_{,y}(1-y\mu_{,y})
-y\mu_{,yy}]-\frac{1}{16}y^4\nu^{-2}\mu_{,u}(\tau_{,y}^2
+\nu_{,y}^2)\nonumber\\
    &&-8\pi\{(1-\frac{1}{2}y\mu_{,y})[S_{,u}^2
+\frac{1}{2}y^3S_{,u}S_{,y}+e^{-2}y^4\nu e^{-2\mu}(P_{,u}^2
+\frac{1}{2}y^3P_{,u}P_{,y})]\nonumber\\
    &&-\frac{1}{8}y^4\mu_{,u}(S_{,y}^2+e^{-2}y^4\nu 
e^{-2\mu}P_{,y}^2) \} \textbf{)} \big/ \bigl\{[y^2-(\mu_{,u}
+\frac{1}{2}\mu_{,y})]^2-\mu_{,u}^2\bigr\}, \\
    \gamma_{,y}&&=\bigl[-\frac{1}{4}(3y^2\mu_{,y}+y^3\mu_{,yy})
+\frac{1}{4}y^3\mu_{,y}^2-\frac{1}{8}y^3\nu^{-2}
(\tau_{,y}^2+\nu_{,y}^2)-2\pi y^3(S_{,y}^2+e^{-2}y^4\nu 
e^{-2\mu}P_{,y}^2)\bigr]\nonumber\\
    &&\big/\bigl(y^2-\frac{1}{2}y^3\mu_{,y}\bigr),
\end{eqnarray}
\begin{eqnarray}
    &&\Box S + \frac{1}{2}y^3S_{,u}\mu_{,y} + \frac{1}{2}y^3 S_{,y}
(\mu_{,u}+\frac{1}{2}y^3\mu_{,y}) =
S[4\lambda\nu^{-1}e^{2(\gamma+\mu)}
(S^2-\eta^2)+y^4e^{2\gamma}P^2],\\
    &&\Box P + P_{,u}[\frac{1}{2}y^3(\nu^{-1}\nu_{,y}-\mu_{,y})+2y^2] 
+ \frac{1}{2}y^3 P_{,y}[\nu^{-1}(\nu_{,u}+\frac{1}{2}y^3\nu_{,y})
-(\mu_{,u}+\frac{1}{2}y^3\mu_{,y})+2y^2] \nonumber\\
    &&= e^2 \nu^{-1}e^{2(\gamma+\mu)} PS^2,\label{compact7}
\end{eqnarray}
where in these coordinates $\Box$ is given by
\begin{equation}
\Box = \frac{y^2}{4} \left(4y\frac{\partial^2}{\partial u\partial y} 
+ y^4 \frac{\partial^2}{\partial y^2} + y^3 \frac{\partial}{\partial
y} - 4\frac{\partial}{\partial u}\right).
\end{equation}
It is worth remarking that one can have solutions to these equations
which are regular at $y=0$ and which represent 4-dimensional metrics in
which $\omega$ diverges. Thus the notion of the 3-dimensional
spacetime being asymptotically flat is weaker than might first be supposed.
The asymptotic behavior of $S$ and $P$ at null infinity is given by
\begin{eqnarray}
  S(u,y) &=& \eta+O(y),\\   
  P(u,y) &=& O(y).
\end{eqnarray}
These are discussed in more detail in section V.

The field equations are solved numerically in two ways. Firstly an
explicit second order Cauchy-Characteristic Matching (CCM) 
scheme similar to that
employed by Dubal et al. \cite{CDD} and d'Inverno et al. \cite{DDS}
is used, but using a Geroch
decomposition in the whole spacetime (not just the characteristic
portion) which allows one to use the geometrically defined variables in
both the interior and exterior regions. This scheme works very well
when compared to the exact vacuum solutions but is less satisfactory
when the matter terms are included (see below). An alternative scheme 
with similar accuracy
but with long term stability is a fully characteristic second order
implicit scheme. This has the advantage that the scheme naturally controls   
the growth of the derivatives at infinity and hence automatically
selects the physical rather than the non-physical solutions. The
details of this scheme are described in paper II.

\section{Exact vacuum solutions}

In this section we describe the exact vacuum solutions which are used
to test the codes. The solutions we will consider are the Weber--Wheeler
gravitational wave \cite{weber} which just has the $+$ polarization mode 
and two solutions due to Xanthopoulos \cite{Xan} and Piran et al.
\cite{Piran} which have both the $+$ and $\times$ polarization mode.

The first exact solution we consider is the Weber--Wheeler 
gravitational wave originally investigated by Einstein and Rosen
\cite{ros1}. It consists of a cylindrically symmetric vacuum wave 
with one radiational degree of freedom corresponding to the $+$ 
polarization mode \cite{PiranII}. It describes a gravitational pulse 
originating from past null infinity and moving toward the $z$-axis. 
After imploding on the axis, it emanates to future null infinity.

This solution has no rotation so that $\omega$ and hence $\tau$
vanish and the solution may be described in terms of $\psi$ which
satisfies the wave equation. A solution to the wave equation in
cylindrical coordinates may be given in terms  of Bessel functions and 
by superposing such solutions we may write 
\begin{equation}
  \psi(t,\rho)=2b\int\limits_0^\infty e^{-a\Omega}
    J_0(\Omega\rho)\cos(\Omega t)d\Omega,\label{D75}
\end{equation}
where $a>0$. For convenience we let
\begin{equation}
  X=a^2+\rho^2-t^2
\end{equation}
and one may show that (\ref{D75}) may be written in the alternative
form
\begin{equation}
  \nu(t,\rho)=
\exp\left[2b \sqrt{\frac{2(X+\sqrt{X^2+4a^2t^2})}
{X^2+4a^2t^2}}\right].
\label{D76}
\end{equation}
The corresponding value of $\gamma$ is obtained by integrating 
$\gamma_{,\rho}=\rho(\psi^2_{,t}+\psi^2_{,\rho})$ and using $\gamma(t,0)=0$
and is found to be
\begin{equation}
  \gamma(t,\rho)=\frac{b^2}{2a^2}\biggl[1-2a^2\rho^2\frac{X^2-4a^2t^2}{(X^2+4a^2t^2)^2} -
    \frac{a^2+t^2-\rho^2}{\sqrt{X^2+4a^2t^2}}\biggr].\label{D77}
\end{equation}

The next solution we consider is one due to Xanthopoulos \cite{Xan}
which has a conical
singularity on the $z$-axis and therefore describes a rotating 
vacuum solution with a cosmic string type singularity.
Xanthopoulos derived the spacetime by finding a
solution to the Ernst equation in prolate spheroidal coordinates. To
compare this with our numerical result we must transform to
cylindrical coordinates and also find the Geroch potential. 
It is convenient to first define the following quantities
\begin{eqnarray}
    &&Q=\rho^2-t^2+1,\\ 
    &&X=\sqrt{Q^2+4t^2},\\
    &&Y=\frac{1}{2}[(2a^2+1)X+Q]+1-a\sqrt{2(X-Q)},\\
    &&Z=\frac{1}{2}[(2a^2+1)X+Q]-1,
\end{eqnarray}
where $0 < |a| < \infty$. The solution derived by Xanthopoulos then becomes
\begin{eqnarray}
    &&\psi(t,\rho)=\frac{1}{2}\ln \frac{Z}{Y},\label{D80}\\
    &&\omega(t,\rho)=\frac{\sqrt{a^2+1}(X+Q-2)(Z-Y)}{2aZ},\label{D81}\\
    &&\gamma(t,\rho)=\frac{1}{2}\ln \frac{Z}{a^2X},\label{D82}
\end{eqnarray}
where we have imposed $\gamma(t,0)=0$. A straightforward but rather tedious 
calculation shows that this satisfies Einstein's field equations. 
(This and a number of other calculations in the paper were checked
using the algebraic computing package GRTensor II \cite{GRII}). 

The norm of the Killing vector in the $z$-direction is given  by
\begin{equation}
  \nu(t,\rho)=\frac{Z}{Y}.
\end{equation}
The Geroch potential is easily obtained from the Ernst potential and
is found to be
\begin{equation}
  \tau(t,\rho)=-\frac{\sqrt{2(a^2+1)}\sqrt{X+Q}}{Y}.\label{xant1}
\end{equation}
Note that $\nu$ and $\tau$ satisfy (\ref{E1})--(\ref{E2}) in vacuum, i.e. 
\begin{eqnarray}
  &&\Box\nu - \nu^{-1}(\tau_{,t}^2-\tau_{,\rho}^2-\nu_{,t}^2+\nu_{,\rho}^2)=0,\\
  &&\Box\tau + 2\nu^{-1}(\tau_{,t}\nu_{,t}-\tau_{,\rho}\nu_{,\rho})=0.
\end{eqnarray}

Expressions for all these quantities may be obtained in the exterior
characteristic region by transforming to the $(u,y)$
variables. Although $\psi$ tends to zero as one approaches null
infinity, $\omega$ diverges so that the 4-dimensional metric is not
asymptotically flat even along null geodesics lying in the planes
$z=\mathrm{const}$. However by
contrast the Geroch potential $\tau$ vanishes as one approaches null
infinity in the 3-dimensional spacetime. This is an example of the
fact that the Geroch potential can
be well behaved even if $\omega$ in the JEKK form of the 4-metric
diverges as one goes outward in a null direction. We also give an
expression for the gravitational flux at infinity which is given by
$E_{,u}$ where $E(u,y)=2\pi[1-e^{-\gamma(u,y)}]$
\begin{equation}
  \lim_{y \to 0}E_{,u}=-\frac{4\pi}{(1+u^2)[(1+2a^2)\sqrt{1+u^2}-u]} < 0.
\end{equation}
Thus the string is losing energy through gravitational radiation. 

To plot the solution for $0\le\rho<\infty$ we introduce the radial variable
\begin{eqnarray}
  w = \left\{ \parbox{50mm} {$\rho \hspace{1.7cm} \, {\rm for} \,\,\, 0\le \rho \le 1\\
                             3-2/{\sqrt\rho} \hspace{0.55cm} {\rm for} \,\,\,
                             \rho>1$ ,}\right.
\end{eqnarray}
thus $0\le w \le 3$ where the infinite value of $\rho$ is mapped to
$w=3$. This choice is slightly different from that of
Dubal et al. \cite{CDD} and avoids discontinuities in the radial
derivatives at the interface due to the square root in the definition of $y$.
Plots of $\nu$, $\tau$ and $\gamma$ as given by (\ref{D82})--(\ref{xant1}) are
shown in Fig. 1. The error in the numerical results as computed using
the CCM code are shown in Fig. 4.

\begin{figure}[ht] \centering
\mbox{{\scalebox{0.31}{\includegraphics {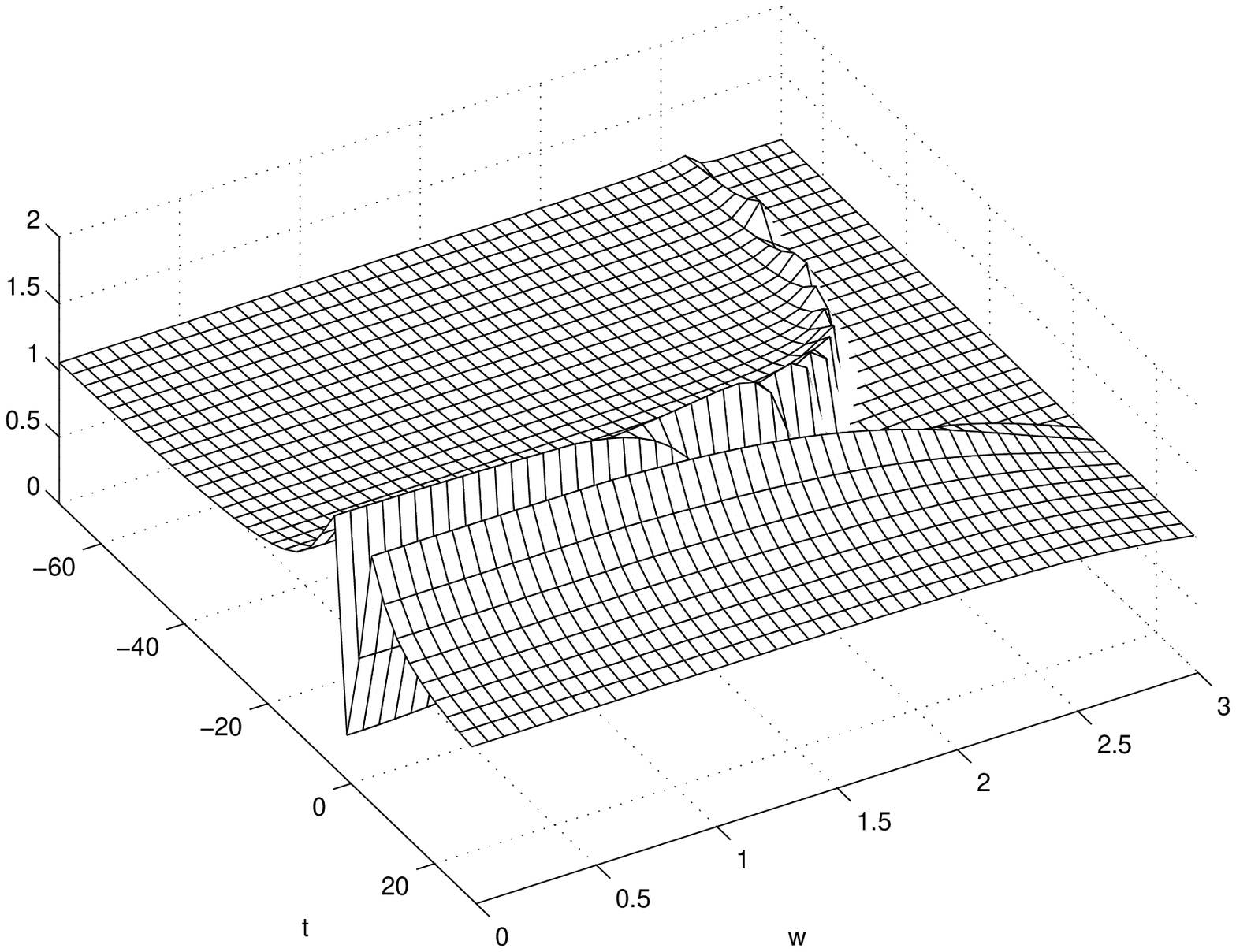}}}
      {\scalebox{0.31}{\includegraphics {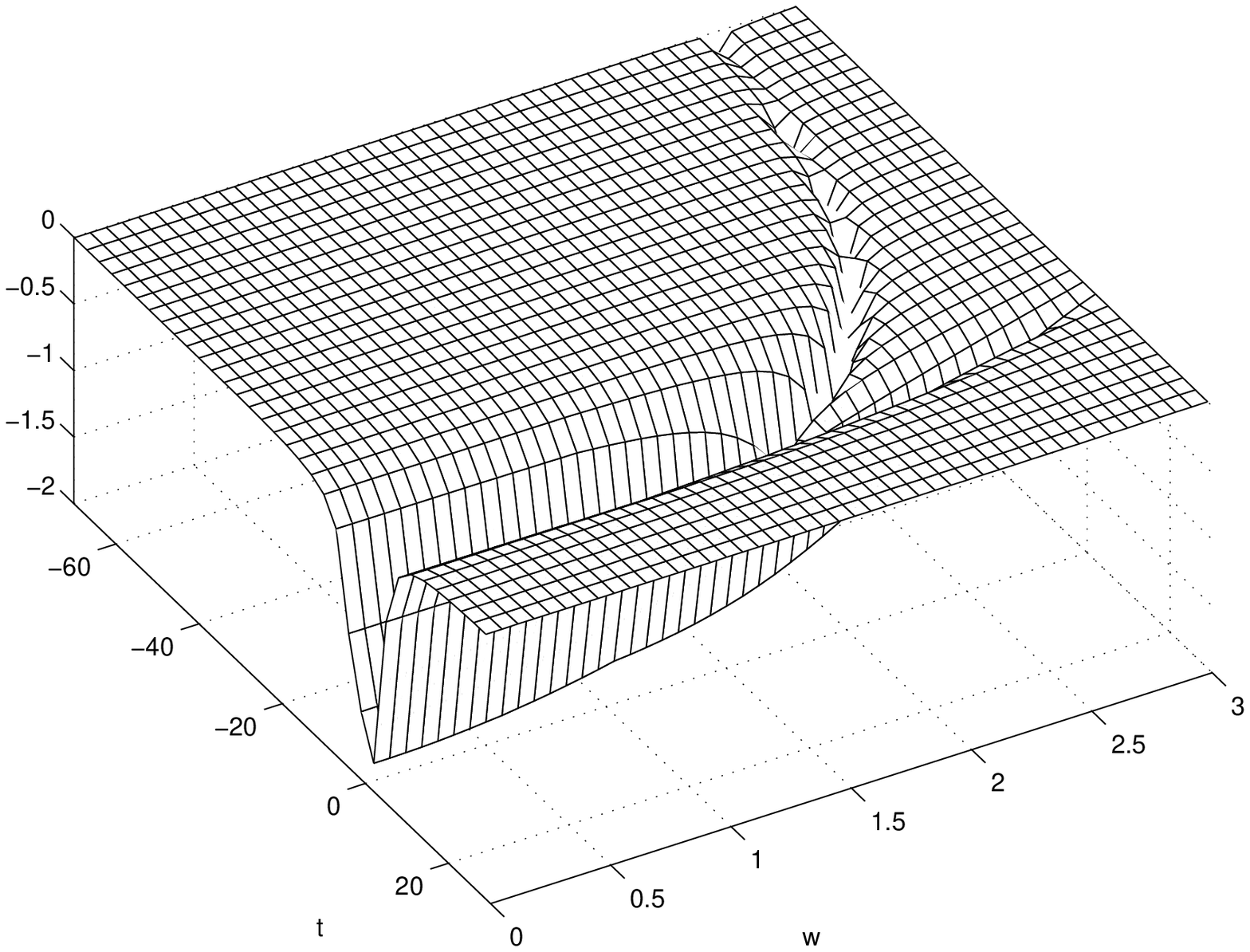}}}
      {\scalebox{0.31}{\includegraphics {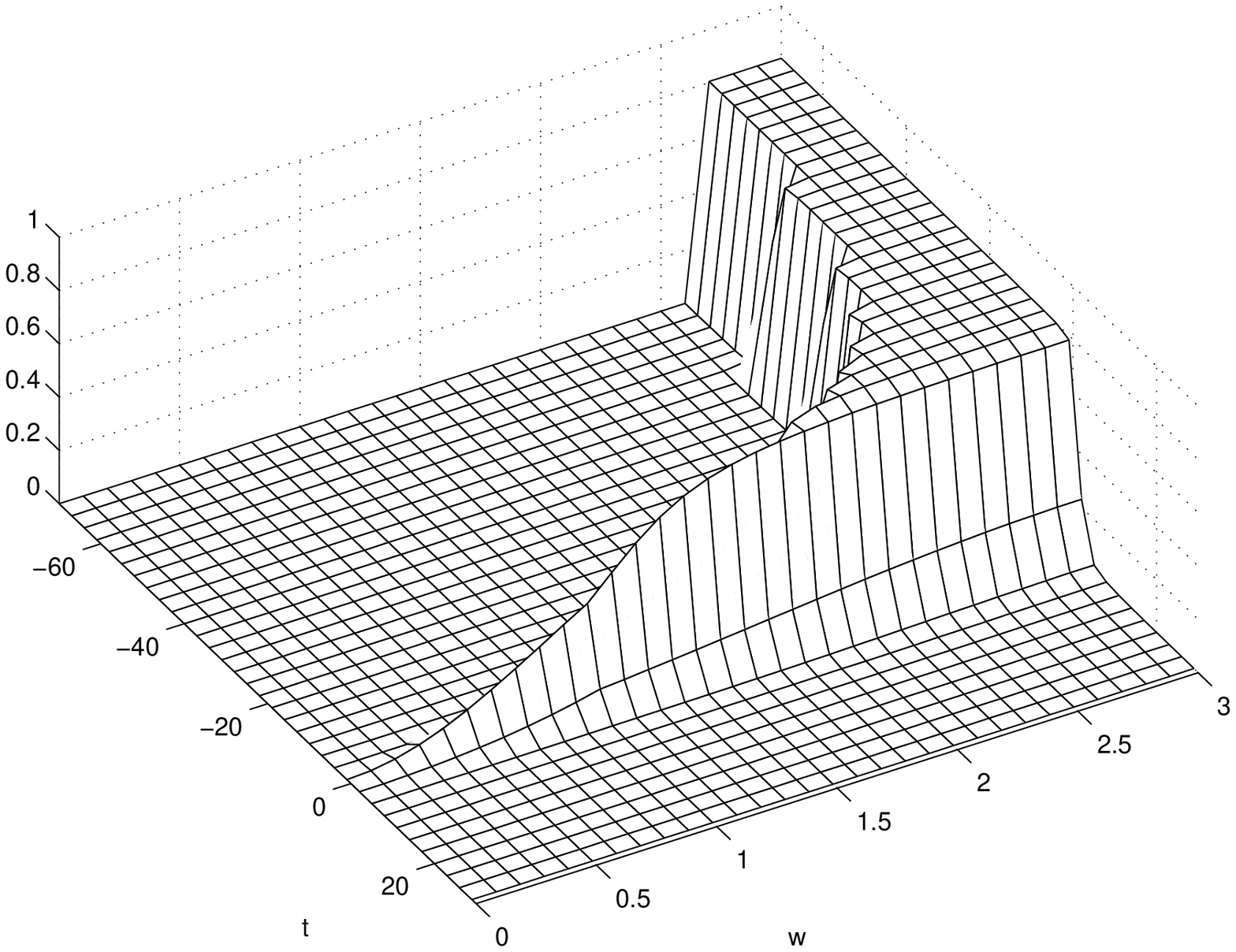}}}}
\caption{The exact Xanthopoulos solution for $-70\le t \le 30$, $0\le w \le
3$ and $a=0.5$. Plots are from left to right: $\nu(t,w)$, $\tau(t,w)$
and $\gamma(t,w)$. One clearly sees the incoming pulses in $\nu$ and
$\tau$. As the pulse hits the axis the string loses energy
through gravitational radiation as seen in $\gamma$.}
\end{figure}

The final exact solution we consider is one due to Piran et al. \cite{Piran} 
which also has two degrees of freedom representing the two
polarization states. As in the case of Xanthopoulos' solution, 
it represents two incoming pulses that implode on the axis and 
then move away from it. Piran et al. obtained their solution by
starting with the Kerr metric in Boyer-Linquist form, transforming to
cylindrical polar coordinates and then swapping the $t$ and $z$
coordinates (and introducing some factors of $i$ to maintain a real
Lorentzian metric). See \cite{Piran} for details. The resulting metric
may be written in JEKK form. The solution is rather
complicated but may be simplified by introducing the following
additional quantities
\begin{eqnarray}
    &&R=b^{-1}[\sqrt{b^2+(t-\rho)^2}-t+\rho],\label{D102a}\\ 
    &&S=b^{-1}[\sqrt{b^2+(t+\rho)^2}+t+\rho],\\
    &&T=1+RS+2a^{-1}\sqrt{(a^2-1)RS}
\end{eqnarray}
and
\begin{eqnarray}
    &&X=(1+R^2)(1+S^2),\\ 
    &&Y=a^2T^2+(R-S)^2,\\
    &&Z=a^2(1-RS)^2+(R+S)^2,\label{D102b}
\end{eqnarray}
where $1 \leq a < \infty$ and $0 \leq b<\infty$. The metric coefficients are
then given by
\begin{eqnarray}
    &&\psi(t,\rho)=\frac{1}{2}\ln \frac{Z}{Y},\label{piran1a}\\
    &&\omega(t,\rho)=b\sqrt{a^2-1}\biggl[2\biggl(1
+\frac{\sqrt{a^2-1}}{a}\biggr)-\frac{(R+S)^2T}{\sqrt{RS}Z}\biggr],\\
    &&\gamma(t,\rho)=\frac{1}{2}\ln \frac{Z}{X}.\label{piran1b}
\end{eqnarray}
Notice that Minkowski space is obtained in the limit that $a \to 1$,
and that we can also consider the case $b \to 0$ in which case the
rotation vanishes. This is not true for the Xanthopoulos solution
which is only real for a sufficiently large rotation.

\begin{figure}[ht] \centering
\mbox{{\scalebox{0.31}{\includegraphics {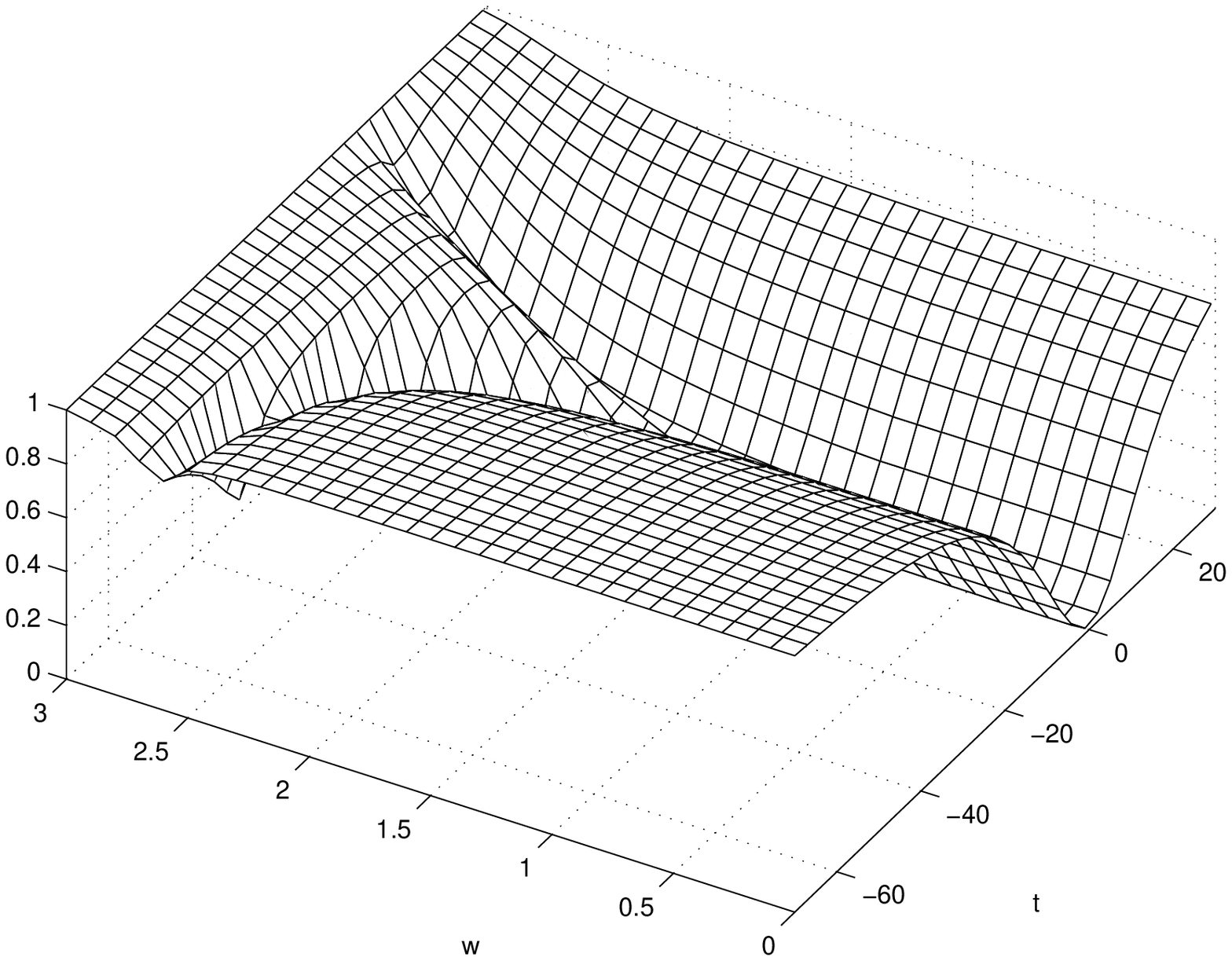}}}
      {\scalebox{0.31}{\includegraphics {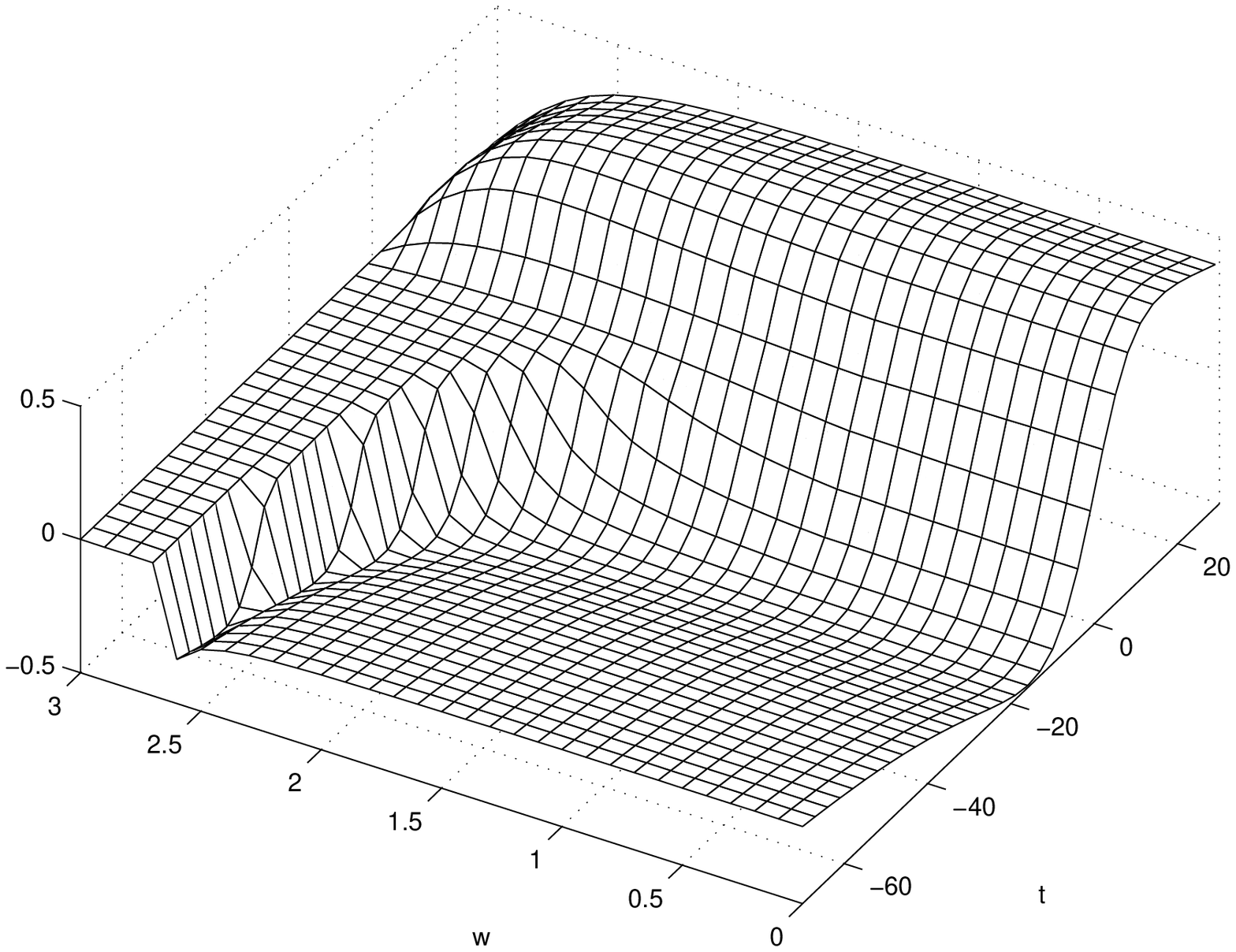}}}
      {\scalebox{0.31}{\includegraphics {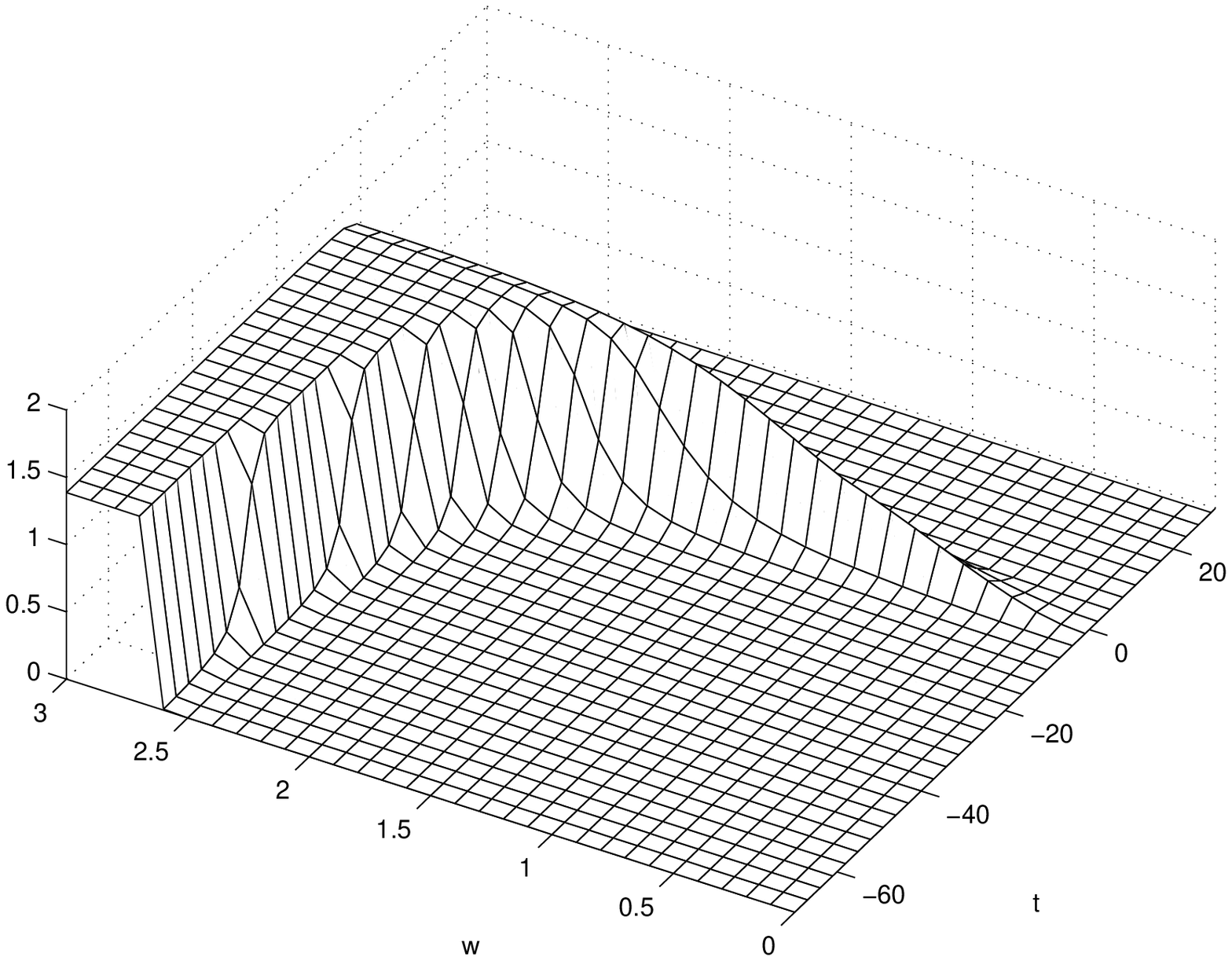}}}}
\caption{The exact Piran et al. solution for $-70\le t \le 30$, $0\le w \le 3$, $a=4$ and $b=2$. Plots are from left to right: $\nu(t,w)$, $\tau(t,w)$ and $\gamma(t,w)$. Notice that $\nu$ does not have a double ridge as found in the Xanthopoulos solution.}
\end{figure}

The solution is regular on the axis, but like the Xanthopoulos
solution $\omega$ diverges as one approaches null infinity. Again the
answer is to transform to the $\nu$, $\tau$ variables which are
regular both on the axis and at null infinity. Finding $\nu$ is
straightforward; however solving the
differential equations for the Geroch potential $\tau$ for such a complicated
metric is extremely difficult, but $\tau$ may be found by first
finding the Geroch potential for the {\it timelike} Killing vector of the
Kerr solution and then making the appropriate transformations. Note
that the same process transforms the Killing vector into one
along the $z$-axis.  One then finds
\begin{eqnarray}
  &&\nu(t,\rho)=\frac{Z}{Y},\\
  &&\tau(t,\rho)=-\frac{4\sqrt{(a^2-1)RS}(R-S)}{[2\sqrt{(a^2-1)RS}
    +a(1+RS)]^2+(R-S)^2}. \label{ptau}
\end{eqnarray}
The corresponding results in the characteristic region are easily
found by transforming to $(u,y)$ coordinates.
Plots of $\nu$, $\tau$ and $\gamma$ as given by
(\ref{piran1b})--(\ref{ptau}) are
shown in Fig. 2. The error in the numerical results as computed using
the CCM code are shown in Fig. 5.

We now briefly describe the accuracy and the convergence analysis for the
explicit CCM version of our code. The results for the the implicit
version are similar and are given in paper II. We define the pointwise 
error at the $i$th time slice and $j$th grid point for some function
$f$ by
\begin{equation}
    \xi_j^i(f) = f(t_i,w_j)_{\mathrm{exact}}-f(t_i,w_j)_{\mathrm{computed}}.
    \end{equation}
The pointwise error for the vacuum solutions  is shown in Fig. 3--5
for 600 grid points and 10,000 time steps corresponding to $0\le t \le 15$. 
The code is stable and accurate for at least 20,000 time steps with a 
Courant factor of $0.45$. Beyond this point the metric functions have almost
 decayed to zero and the dynamical behaviour is very slow.
\begin{figure}[ht] \centering
\mbox{{\scalebox{0.31}{\includegraphics {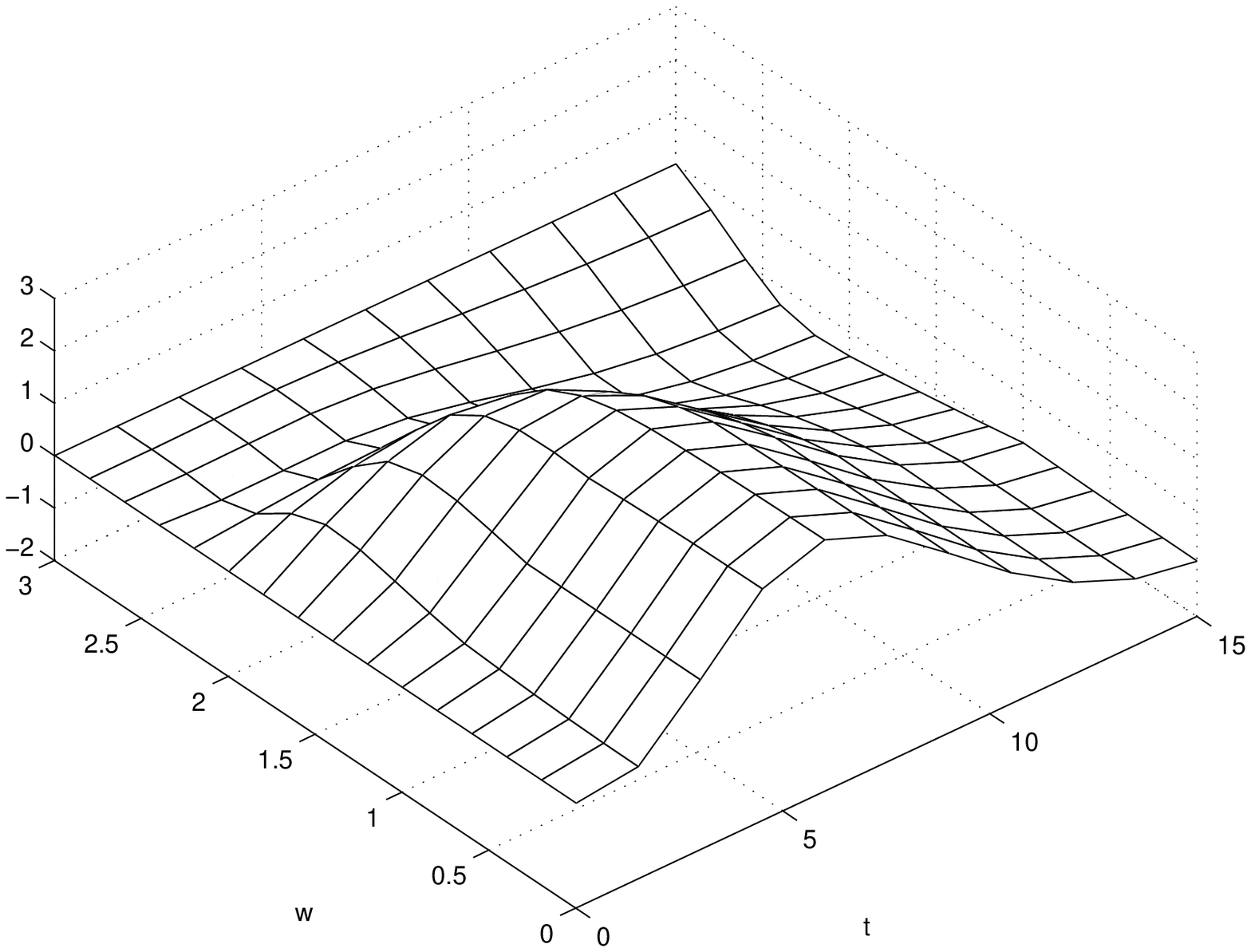}}}
      {\scalebox{0.31}{\includegraphics {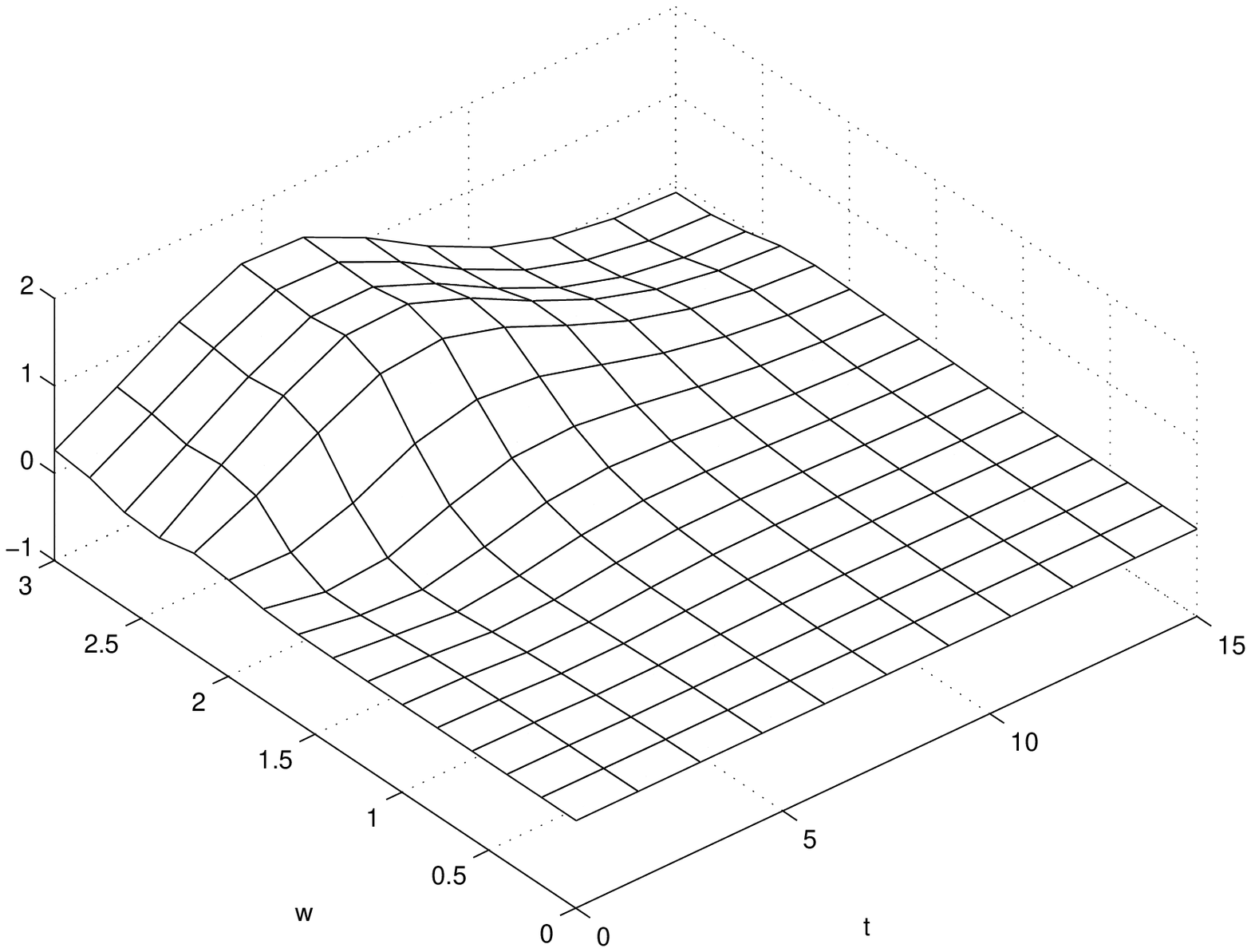}}}}
\caption{Pointwise error for the Weber--Wheeler solution. From left to right: $\psi(t,w)\cdot 10^6$ and $\gamma(t,w)\cdot 10^7$. }
\end{figure}
\begin{figure}[ht] \centering
\mbox{{\scalebox{0.31}{\includegraphics {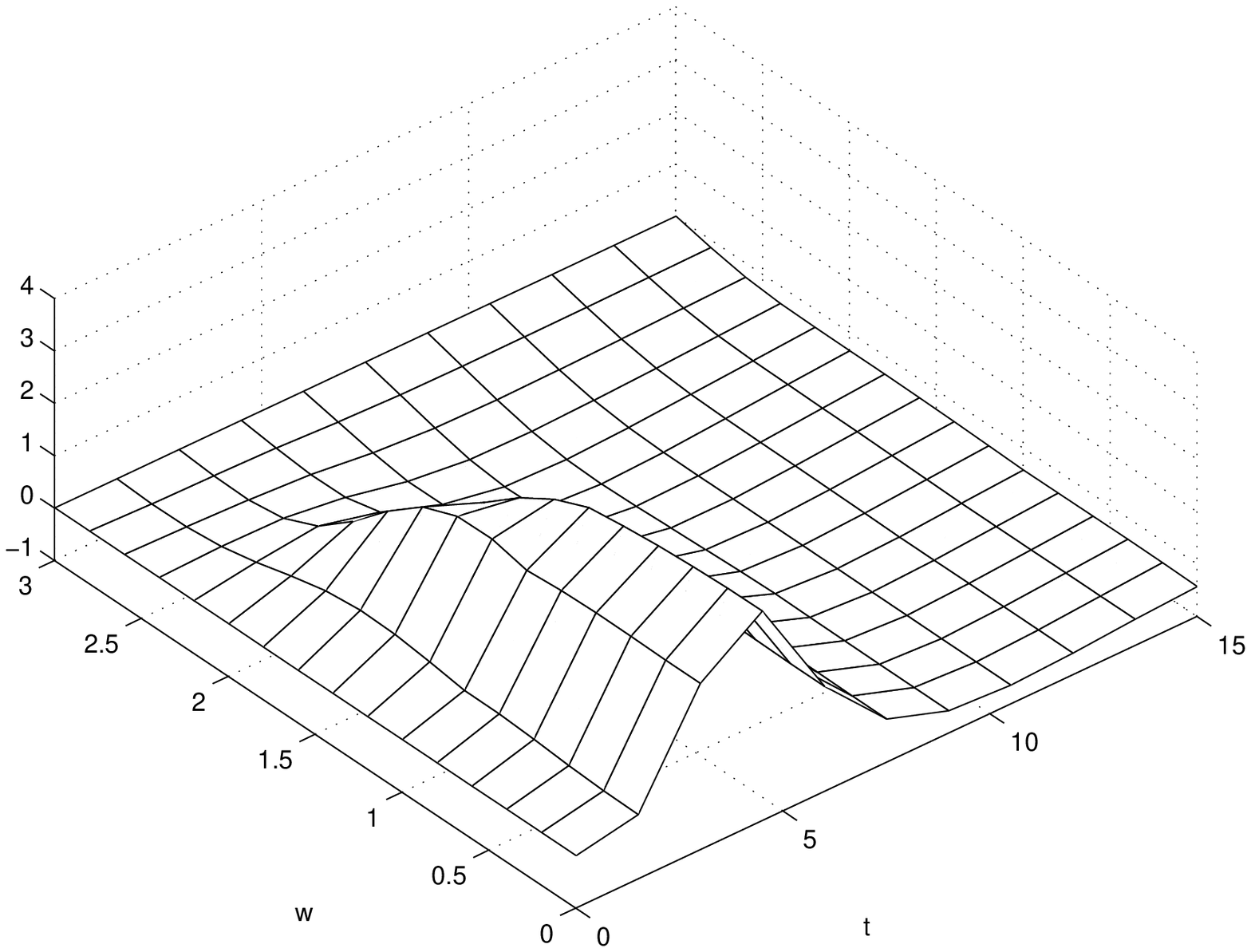}}}
      {\scalebox{0.31}{\includegraphics {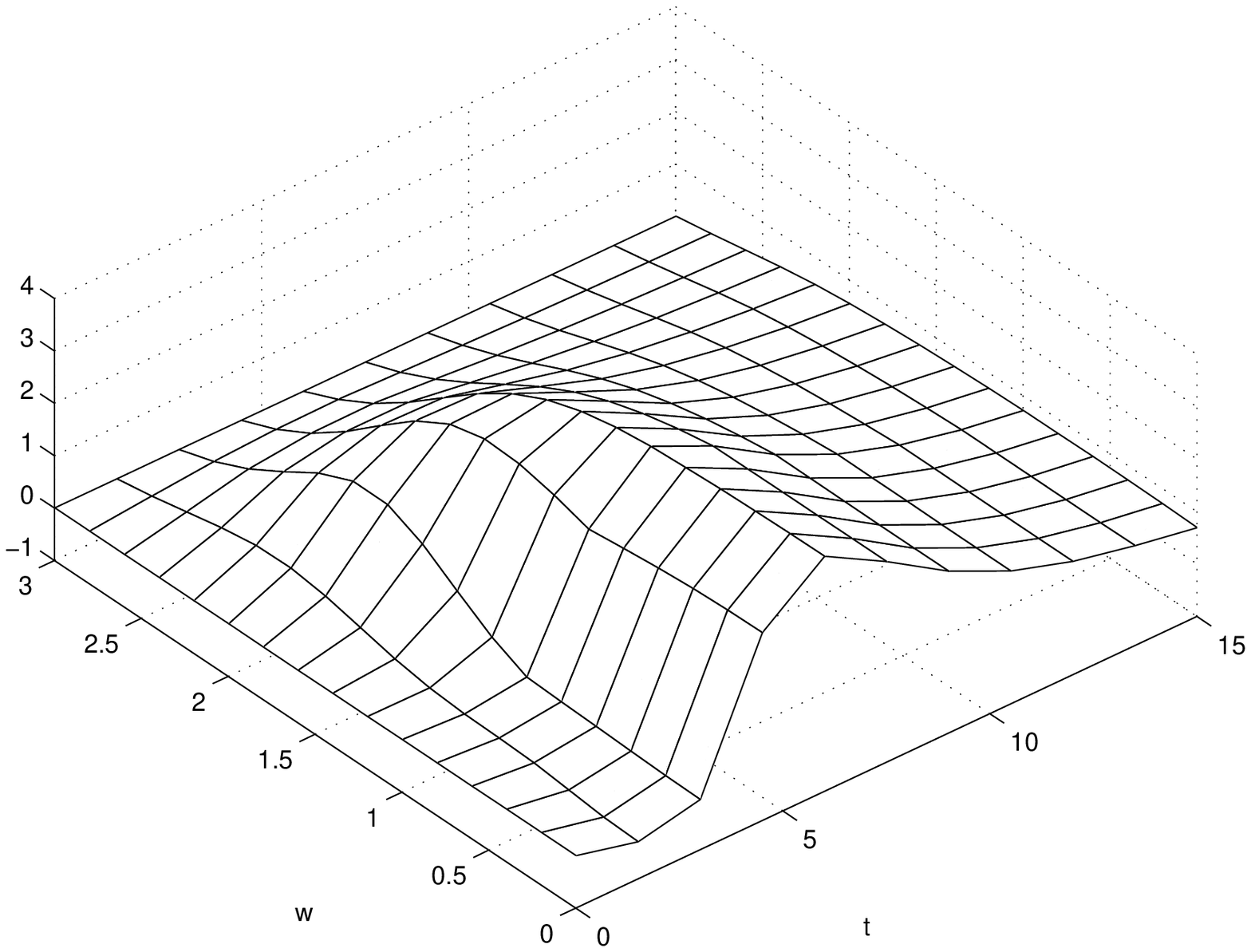}}}}
\caption{Pointwise error for the Xanthopoulos solution. From left to
right: $\nu(t,w)\cdot 10^5$ and $\tau(t,w)\cdot 10^5$.}
\end{figure}
\begin{figure}[ht] \centering
\mbox{{\scalebox{0.31}{\includegraphics {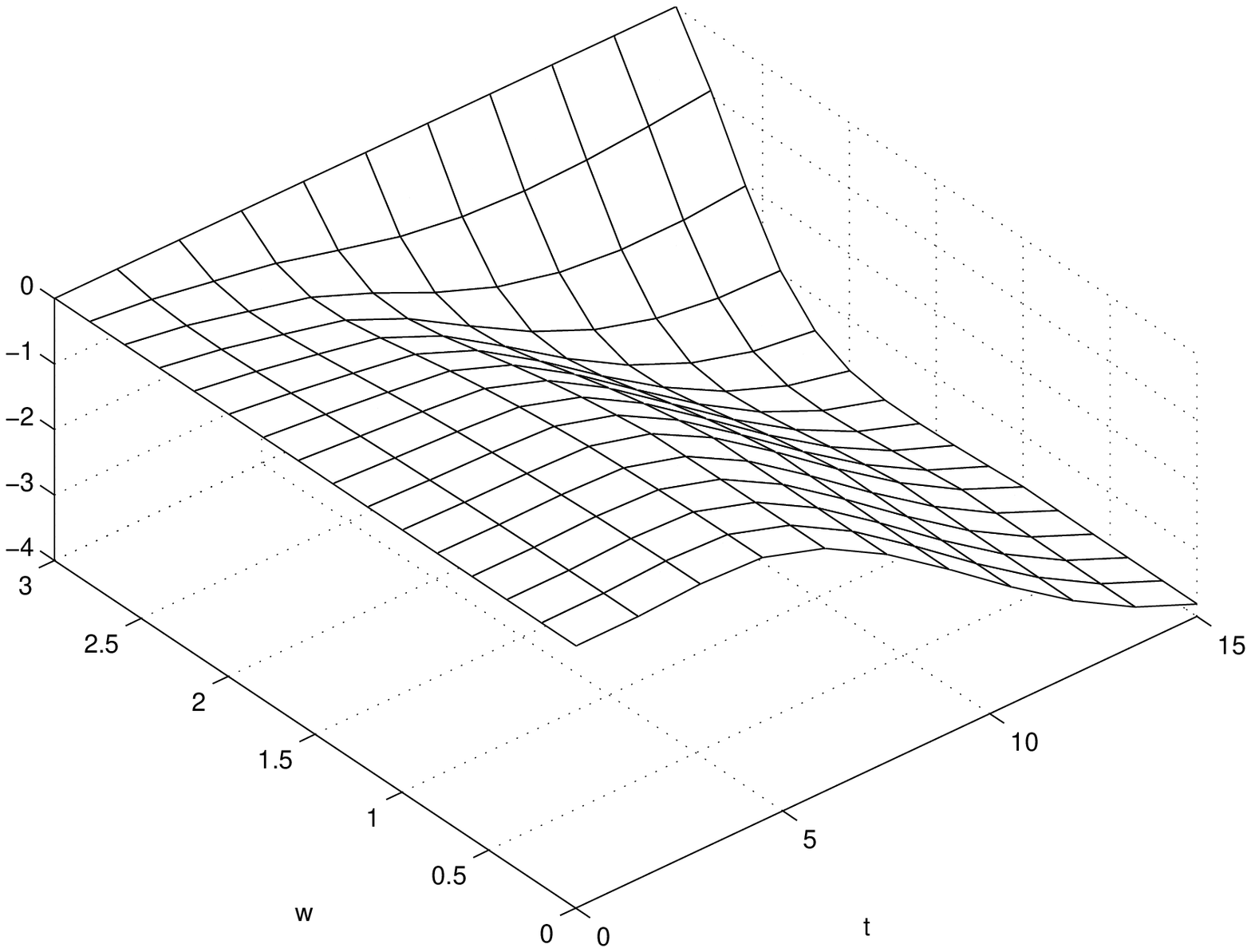}}}
      {\scalebox{0.31}{\includegraphics {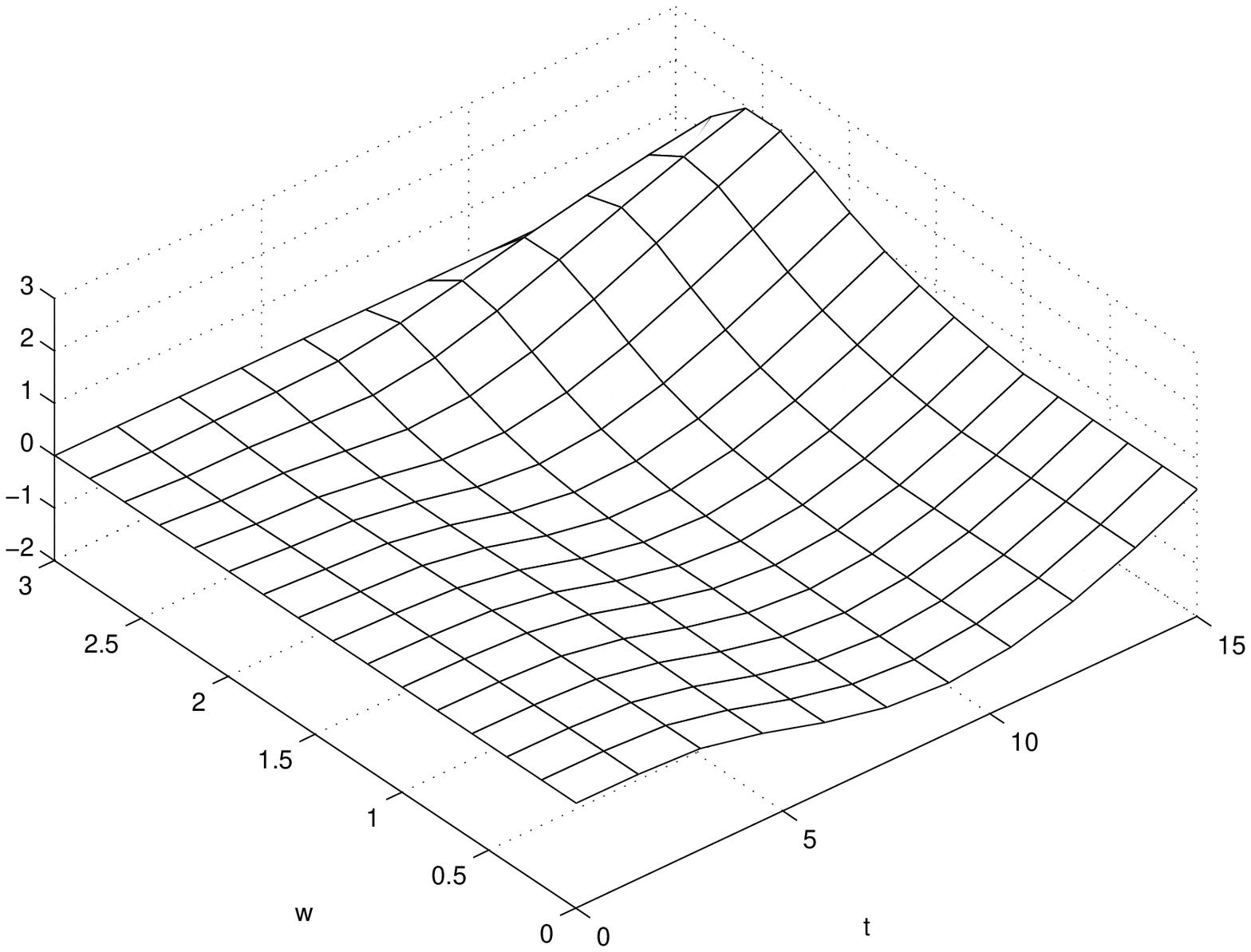}}}}
\caption{Pointwise error for the Piran et al. solution. From left to
right: $\nu(t,w)\cdot 10^5$ and $\tau(t,w)\cdot 10^5$.}
\end{figure}

In order to analyze the convergence of the code we define the spacetime $\ell_2$-norm for some function $f$ as 
    \begin{equation}
    \ell_2[f^{N_1}_{N_2}] = \sqrt{\frac{\sum_{i,j} [\xi_j^i(f)]^2}{N_1\cdot N_2}},
    \end{equation}
where $N_1$ is the number of time slices and $N_2$ is the number of
grid points on each slice. We also define the relative norm
    \begin{equation}
    f_r = \sqrt{\frac{\sum_{i,j}[\xi_j^i(f)]^2}
          {\sum_{i,j} [f^i_{j\ \mathrm{exact}}]^2}}.
    \end{equation}
Convergence testing of the code is done by doubling the grid size keeping
the Courant factor constant. We therefore also need to double the
number of time steps. We measure the convergence through
    \begin{equation}
      f_c = \frac{\ell_2[f^{N_1}_{N_2}]}{\ell_2[f^{2N_1}_{2N_2}]}.
    \end{equation}
For a convergent second order code one should have $f_c=4$, and we can
clearly see from Tables I--III that the code has achieved second order 
convergence in time and space.
    \begin{center}
    \begin{tabular*}{102.6mm}{|c|c|c|c|c|c|}
    \hline\hline
    \multicolumn{1}{|c|}{\,\, Grid pts.\,\,} & \multicolumn{1}{c|}{ $\nu_r$ } & \multicolumn{1}{c|}{ $\gamma_r$ }& \multicolumn{1}{c|}{\,\,\,Time steps\,\, } & \multicolumn{1}{c|}{ $\nu_c$ } & \multicolumn{1}{c|}{ $\gamma_c$ }  \\
    \hline
    300   & $\,\, 6.29\cdot 10^{-7}\,\,$  & $\,\, 9.38\cdot 10^{-6} \,\,$ & 5,000   & $\,\, -\,\,$ & $\,\, -\,\,$ \\
    600   & $\,\, 1.11\cdot 10^{-7}\,\,$  & $\,\, 1.66\cdot 10^{-6} \,\,$ & 10,000  & \,\,4.01 \,\,&\,\, 4.01\,\, \\
    1,200 & $\,\, 1.96\cdot 10^{-8}\,\,$  & $\,\, 2.94\cdot 10^{-7} \,\,$ & 20,000  & \,\,4.00 \,\,&\,\, 4.00\,\, \\
    2,400 & $\,\, 3.47\cdot 10^{-9}\,\,$  & $\,\, 5.19\cdot 10^{-8} \,\,$ & 40,000  & \,\,4.00 \,\,&\,\, 4.00\,\, \\
    4,800 & $\,\, 6.14\cdot 10^{-10}\,\,$ & $\,\, 9.18\cdot 10^{-9} \,\,$ & 80,000  & \,\,4.00 \,\,&\,\, 4.00\,\, \\
    \hline\hline
    \multicolumn{6}{c}{\parbox{82mm}{\small{TABLE I. Convergence test:
    the Weber--Wheeler solution.}}}
    \end{tabular*}
    \end{center}

    \begin{center}
    \begin{tabular*}{111.8mm}{|c|c|c|c|c|c|c|}
    \hline\hline
    \multicolumn{1}{|c|}{\,\, Grid pts.\,\, } & \multicolumn{1}{c|}{ $\nu_r$ } & \multicolumn{1}{c|}{ $\tau_r$ } & \multicolumn{1}{c|}{ $\gamma_r$ } & \multicolumn{1}{c|}{ $\nu_c$ } & \multicolumn{1}{c|}{ $\tau_c$ } & \multicolumn{1}{c|}{ $\gamma_c$ }  \\
    \hline
    300   & $\,\, 1.70\cdot 10^{-6}\,\,$ & $\,\, 5.71\cdot 10^{-6}\,\,$ & $\,\, 1.66\cdot 10^{-6}\,\,$ & $\,\,-\,\,$  & $-$  & $-$  \\
    600   & $\,\, 2.99\cdot 10^{-7}\,\,$ & $\,\, 1.01\cdot 10^{-6}\,\,$ & $\,\, 2.93\cdot 10^{-7}\,\,$ &\,\, 4.01\,\, &\,\, 4.00\,\, &\,\, 4.01\,\, \\
    1,200 & $\,\, 5.28\cdot 10^{-8}\,\,$ & $\,\, 1.78\cdot 10^{-7}\,\,$ & $\,\, 5.18\cdot 10^{-8}\,\,$ &\,\, 4.01\,\, &\,\, 4.00\,\, &\,\, 4.01\,\, \\
    2,400 & $\,\, 9.32\cdot 10^{-9}\,\,$ & $\,\, 3.15\cdot 10^{-8}\,\,$ & $\,\, 9.16\cdot 10^{-9}\,\,$ &\,\, 4.00\,\, &\,\, 4.00\,\, &\,\, 4.00\,\, \\
    4,800 & $\,\, 1.65\cdot 10^{-9}\,\,$ & $\,\, 5.57\cdot 10^{-9}\,\,$ & $\,\, 1.62\cdot 10^{-9}\,\,$ &\,\, 4.00\,\, &\,\, 4.00\,\, &\,\, 4.00\,\, \\
    \hline\hline
    \multicolumn{7}{c}{\parbox{80mm}{\small{TABLE II. Convergence test:
    the Xanthopoulos solution.}}}
    \end{tabular*}
    \end{center}

    \begin{center}
    \begin{tabular*}{111.8mm}{|c|c|c|c|c|c|c|}
    \hline\hline
    \multicolumn{1}{|c|}{\,\, Grid pts.\,\, } & \multicolumn{1}{c|}{ $\nu_r$ } & \multicolumn{1}{c|}{ $\tau_r$ } & \multicolumn{1}{c|}{ $\gamma_r$ } & \multicolumn{1}{c|}{ $\nu_c$ } & \multicolumn{1}{c|}{ $\tau_c$ } & \multicolumn{1}{c|}{ $\gamma_c$ }  \\
    \hline
    300   & $\,\, 1.07\cdot 10^{-5}\,\,$ & $\,\, 3.94\cdot 10^{-6}\,\,$ & $\,\, 2.76\cdot 10^{-6}\,\,$ & $\,\, -\,\, $  & $\,\, -\,\, $  & $\,\, -\,\, $  \\
    600   & $\,\, 1.89\cdot 10^{-6}\,\,$ & $\,\, 6.96\cdot 10^{-7}\,\,$ & $\,\, 4.87\cdot 10^{-7}\,\,$ &\,\, 4.00\,\, &\,\, 4.00\,\, &\,\, 4.01\,\, \\
    1,200 & $\,\, 3.35\cdot 10^{-7}\,\,$ & $\,\, 1.23\cdot 10^{-7}\,\,$ & $\,\, 8.61\cdot 10^{-8}\,\,$ &\,\, 4.00\,\, &\,\, 4.00\,\, &\,\, 4.00\,\, \\
    2,400 & $\,\, 5.92\cdot 10^{-8}\,\,$ & $\,\, 2.17\cdot 10^{-8}\,\,$ & $\,\, 1.52\cdot 10^{-8}\,\,$ &\,\, 4.00\,\, &\,\, 4.00\,\, &\,\, 4.00\,\, \\
    4,800 & $\,\, 1.05\cdot 10^{-8}\,\,$ & $\,\, 3.84\cdot 10^{-9}\,\,$ & $\,\, 2.69\cdot 10^{-9}\,\,$ &\,\, 4.00\,\, &\,\, 4.00\,\, &\,\, 4.00\,\, \\
    \hline\hline
    \multicolumn{6}{c}{\parbox{78mm}{\small{TABLE III. Convergence
    test: the Piran et al. solution.}}}
    \end{tabular*}
    \end{center}

\section{The Cosmic String in Minkowski spacetime}

In this section we examine the field equations for the cosmic
string. The simplest case is to look at the equations of motion on a
fixed Minkowski background. If we do this then the Euler--Lagrange
equations for (\ref{1.1}) give
\begin{eqnarray}
  &&\Box S=S[4\lambda(S^2-\eta^2)-\rho^{-2}P^2], \label{a} \\
  &&\Box P-2\rho^{-1}P_{,\rho}=e^2S^2P, \label{b}
\end{eqnarray}
where $\Box$ is the d'Alembertian in cylindrical polar coordinates
given by (\ref{dal}).

It will turn out that the solutions of this simpler set of equations
are qualitatively similar to those of the full system of equations for a
dynamic cosmic string coupled to Einstein's equations. However the full system
enables one to perturb a static string with a pulse of gravitational
radiation and in turn look at the effect of the string's oscillations
on the gravitational waves.

A special case of (\ref{a}),  (\ref{b}) is when one looks for a static
solution. This has been looked at before by a number of authors, for
example Garfinkle \cite{garfinkle}, Laguna et al. \cite{sstringI}, 
Laguna--Castillo et al. \cite{sstringII} and Dyer et al. \cite{sstringIII}. 
For a static string
equations (\ref{a}) and (\ref{b}) reduce to
\begin{eqnarray}
    &&\rho\frac{d}{d\rho}\biggl(\rho\frac{dS}{d\rho}\biggr) = 
      S[4\lambda\rho^2(S^2-\eta^2)+P^2], \label{S7}\\
    &&\rho\frac{d}{d\rho}\biggl(\rho^{-1}\frac{dP}{d\rho}\biggr) = 
      e^2S^2P. \label{S8}
\end{eqnarray}
This pair of coupled second order equations requires four boundary
conditions. For the physically relevant finite energy solution these are
\begin{eqnarray}
  &S(0)=0, \qquad\qquad &\lim_{\rho \to \infty} S(\rho)=\eta,\\ 
  &P(0)=1, \qquad\qquad &\lim_{\rho \to \infty} P(\rho)=0.
\end{eqnarray}

It is not possible to obtain an exact solution to these equations but
one can investigate the asymptotic behavior for large $\rho$. 
The solution satisfying the above boundary conditions has asymptotic
behaviour given by \cite{shellard}
\begin{eqnarray}
    &&S(\rho)\sim \eta-K_0(\sqrt{8\lambda}\eta\rho)\sim\eta
-\sqrt{\frac{\pi}{\sqrt{32\lambda}\eta}}\rho^{-1/2}
e^{-\sqrt{8\lambda}\eta\rho},\\
    &&P(\rho)\sim \rho K_1(e\eta\rho)\sim\sqrt{\frac{\pi}{2e\eta}}
\rho^{1/2}e^{-e\eta\rho}.
\end{eqnarray}
However as well as these physical solutions, the equations admit
non-physical solutions which have exponentially divergent behavior as
$\rho \to \infty$. It is the existence of these non-physical solutions
which make the problem rather delicate from a numerical point of view
and makes a method such as shooting hard to apply. 

Before proceeding further we follow Garfinkle \cite{garfinkle} by
introducing rescaled variables and constants which simplify
the algebra (and are also important when considering the thin string
limit). Provided we rescale the time coordinate this also
simplifies the fully coupled system. Let us introduce  
\begin{eqnarray}
    X &=& \frac{S}{\eta},\\
    r &=& \sqrt\lambda \eta\rho, \\
    {\tilde t} &=& \sqrt\lambda \eta t,\\ 
    \alpha &=& \frac{e^2}{\lambda}.
\end{eqnarray}
Thus $\alpha$ represents the relative strength of the coupling between
the scalar and vector field given by $e$ compared to the self-coupling
of the scalar field given by $\lambda$. Critical coupling, when the
masses of the scalar and vector fields are equal, is given by
$\alpha=8$. In the theory of superconductivity, $\alpha=8$ corresponds
to the interface between type I and type II behaviour \cite{shellard}.

With the above rescaling equations (\ref{S7}) and (\ref{S8}) become
\begin{eqnarray}
  &&r\frac{d}{dr}\left( r\frac{dX}{dr} \right) = X[4r^2 (X^2-1) + P^2], \label{6}\\
  &&r\frac{d}{dr} \left( r^{-1} \frac{dP}{dr} \right) = \alpha X^2 P. \label{7}
\end{eqnarray}

Note the rescaled version of (\ref{E1})--(\ref{E1a}) 
may be found simply by making
the replacements
\begin{eqnarray}
  S &\to& X, \\
  \rho &\to& r, \\
  t &\to& {\tilde t},\\
  e^2 &\to& \alpha, \\
  \lambda &\to& 1.
\end{eqnarray}

The boundary conditions of the cosmic string are given by
\begin{eqnarray}
  &X(0) = 0, \qquad \qquad &\lim_{r \rightarrow \infty} X(r) = 1,\\
  &P(0) = 1,  \qquad \qquad &\lim_{r\rightarrow \infty} P(r) = 0.
\end{eqnarray}

Because of the boundary conditions at infinity it  is desirable 
to introduce a new coordinate which brings in infinity to a finite 
coordinate value. For this purpose we again introduce an inner region
($r \le 1$) and an outer region ($r \ge 1$) where we use the radial
coordinate $y$ given by
\begin{equation}
  y = \frac{1}{\sqrt{r}}. 
\end{equation}

In order to solve equations (\ref{6}), (\ref{7}) numerically we
introduce a spatial grid consisting of $n_1$ points in the inner region
and $n_2$ points in the outer region. 
The points $r_1, \ldots, r_{n_1}$ cover the range $0 \le r \le 1$, and
$y_{n_1+1}, \ldots, y_{n_1+n_2}$ cover the range $ 1 \ge y \ge 0$. 
Thus, $r=1=y$ is represented by two points. 
These two points form the interface of the code
where $r$-derivatives are transformed into $y$-derivatives. The static
equations in the outer region are
\begin{eqnarray}
  &&y\frac{d}{dy} \left( y \frac{dX}{dy} \right)  = 
  4X\left[4\frac{(X^2-1)}{y^4}+ P^2 \right],\\
  &&y\frac{d}{dy} \left( y^5\frac{dP}{dy} \right) = 4\alpha X^2 P.
\end{eqnarray}

In order to apply boundary conditions at both $r=0$ and $y=0$ the
equations were solved numerically using a relaxation scheme (as
described in \cite{relax} for example).
For this purpose we wrote the equations as a first order 
system in both regions (see paper II) and used second order centered finite 
differencing. Solutions were computed in this way for different 
resolutions $N=n_1+n_2$ to check the 
convergence of the code. Since there is no exact solution
available the convergence was checked by calculating the $\ell_2$-norm 
with respect to a high resolution result for $N=2400$ 
(1200 points in each region)
\begin{equation}
  \ell_2[\Delta f^N] = \sqrt{\frac{\sum_{i=1}^{N} (f^N_i-f^{2400}_i)^2} {N}},
\end{equation}
where $f$ stands for either $X$ or $P$. For a second order
convergent code one would expect the $\ell_2$-norm to decrease 
by a factor of 4 if the number of grid points is doubled. We find that the
code shows clear second order convergence. Since the convergence in
this case is very similar to that of a static string coupled to
gravity considered in the next section, we only give the results for
the latter case in Table IV.

\begin{figure}[ht]\centering
\scalebox{0.6}[0.5]{\includegraphics {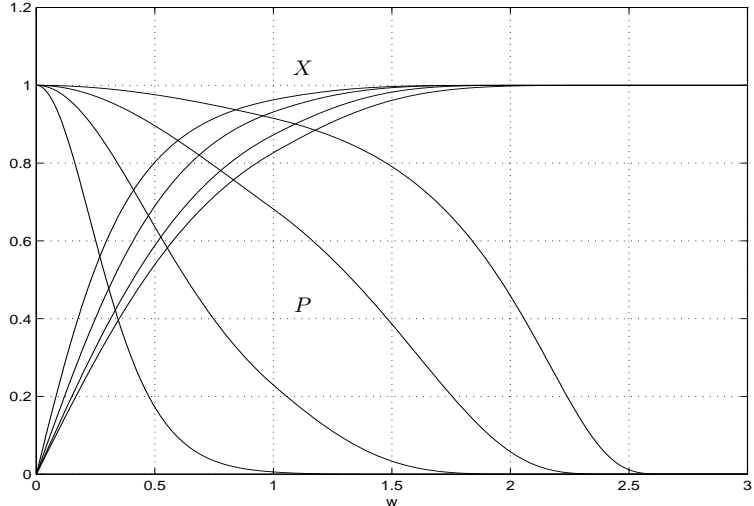}}
\put(-172,170){\fontsize{8}{8pt}\selectfont\makebox(0,0)[t]{$X$}}
\put(-172,80){\fontsize{8}{8pt}\selectfont\makebox(0,0)[t]{$P$}}
\caption{Cosmic string in Minkowski spacetime. $X$ and $P$ for $\alpha$: 0.125, 1, 8, 64.  
As $\alpha$ increases, both $X$ and $P$ become more concentrated towards the origin.}
\end{figure}

In Fig. 6. we plot $X$ and $P$ for $\alpha=0.125,\, 1,\, 8,\, 64$. The
results show that for fixed values of the self-coupling $\lambda$ and
vacuum expectation value $\eta$ of the scalar field, both the vector
and scalar fields become more concentrated towards the origin as the
coupling between the fields $e$ increases. One also finds that for a
fixed ratio of the coupling constants $\alpha$ the fields become more
concentrated towards the origin as either the self-coupling $\lambda$
or the vacuum expectation value $\eta$ of the scalar field increases.

\section{The static Cosmic String coupled to gravity}

The next class of solutions we wish to consider are static solutions
of the fully coupled equations. In the case of no $t$ dependence
equations (\ref{E1})--(\ref{E1a}) reduce to 
\begin{eqnarray}
  \frac{1}{r}(r\nu_{,r})_{,r} &&= -\nu_{,r} \mu_{,r} + \frac{\nu_{,r}^2
                       - \tau_{,r}^2}{\nu} + 8\pi\eta^2\left[
                       \nu^2 e^{-2\mu} \frac{P_{,r}^2}{\alpha r^2}
                       - 2e^{2(\gamma+\mu)}(X^2-1)^2\right], \\
  \frac{1}{r}(r\tau_{,r})_{,r} &&= \tau_{,r}\left(2\frac{\nu_{,r}}{\nu}-
   \mu_{,r}\right)\ , \label{taueq}\\
  \frac{1}{r^2}(r^2\mu_{,r})_{,r} &&= -\mu_{,r}^2 - 8\pi \eta^2 \left[ e^{2\gamma}
              \frac{X^2 P^2}{r^2} + 2 \frac{ e^{2(\gamma+\mu)}}{\nu}(X^2-1)^2\right], \\
  \gamma_{,r} &&=\frac{r}{1+r\mu_{,r}} \left\{
              \frac{1}{4\nu^2}\left(\tau_{,r}^2 + \nu_{,r}^2\right)
              + 4\pi \eta^2 \left[ X_{,r}^2
              + \nu e^{-2\mu} \frac{P_{,r}^2}{\alpha r^2}
              - e^{2\gamma} \frac{X^2 P^2}{r^2} - 2\frac{e^{2(\gamma+\mu)}}{\nu}
              (X^2-1)^2 \right] \right\} -\mu_{,r}\ , \\
  \frac{1}{r}(rX_{,r})_{,r} &&= -X_{,r} \mu_{,r}
              +X\left[
             4 \frac{e^{2(\gamma+\mu)}}{\nu} (X^2-1)+ e^{2\gamma} \frac{P^2}{r^2}\right]\ ,\\
  r\left(\frac{1}{r}P_{,r}\right)_{,r}
              &&= P_{,r}\left(\mu_{,r} - \frac{\nu_{,r}}{\nu}\right)
            +\alpha \frac{ e^{2(\gamma+\mu)}}{\nu} PX^2.
\end{eqnarray}
Notice that (\ref{taueq}) with the corresponding boundary 
conditions is satisfied by the trivial solution $\tau=0$. One also has from 
the field equations
\begin{equation}
  (r\gamma_{,r})_{,r} = -r\gamma_{,r} \mu_{,r} + \mu_{,r}
              + 8\pi \eta^2 \left(e^{2\gamma} \frac{X^2 P^2}{r} +
              e^{-2\mu} \nu \frac{P_{,r}^2}{\alpha r}\right).
  \label{gammarr}
\end{equation}
This equation is a direct consequence of the other equations and will
not be used in the calculations but is instead used as a check for the code.
The equations in the outer region as well as the resulting first order system,
the interface equations and boundary conditions are given in paper II.
We use the same grid and numerical method as in the Minkowskian case. 
In order to check the code for convergence we again compute the 
$\ell_2$-norm with respect to a high resolution 
calculation. The results are shown in Table IV for $\alpha=1$ and the
large value $\eta=0.1$ and clearly
indicate second order convergence. Small deviations from a convergence 
factor of 4 are expected since we compare against a high resolution 
reference solution rather than an exact solution.
The same result has been obtained for other
choices of $\alpha$ and $\eta$. 

  \begin{center}
  \begin{tabular*}{116.4mm}{|c|c|c|c|c|c|}
  \hline\hline
    \multicolumn{1}{|c|}{ } & \multicolumn{1}{c|}{ $\nu$ } & \multicolumn{1}{c|}{ $\mu$ } & \multicolumn{1}{c|}{ $\gamma$ } & \multicolumn{1}{c|}{ $X$ } & \multicolumn{1}{c|}{ $P$ }  \\
      \hline
      $\ell_2(f^{1200})$ & $1.28\cdot 10^{-7}$ & $2.51\cdot 10^{-6}$
                        & $2.39\cdot 10^{-6}$ & $4.16\cdot 10^{-7}$
                        & $5.95\cdot 10^{-7}$ \\
      \hline
      $\ell_2(f^{150})/\ell_2(f^{300})$
                        & 3.56 & 3.59 & 3.58 & 3.37 & 4.04 \\
      $\ell_2(f^{300})/\ell_2(f^{600})$
                        & 3.76 & 3.79 & 3.78 & 3.60 & 4.19 \\
      $\ell_2(f^{600})/\ell_2(f^{1200})$
                        & 4.58 & 4.61 & 4.60 & 4.44 & 4.98 \\
    \hline\hline
    \multicolumn{6}{c}{\parbox{101mm}{\small{TABLE IV. Convergence
  test: static cosmic string coupled to gravity.}}}
    \end{tabular*}
     \end{center}

\begin{figure}[ht] \centering
\mbox{{\scalebox{0.45}{\includegraphics {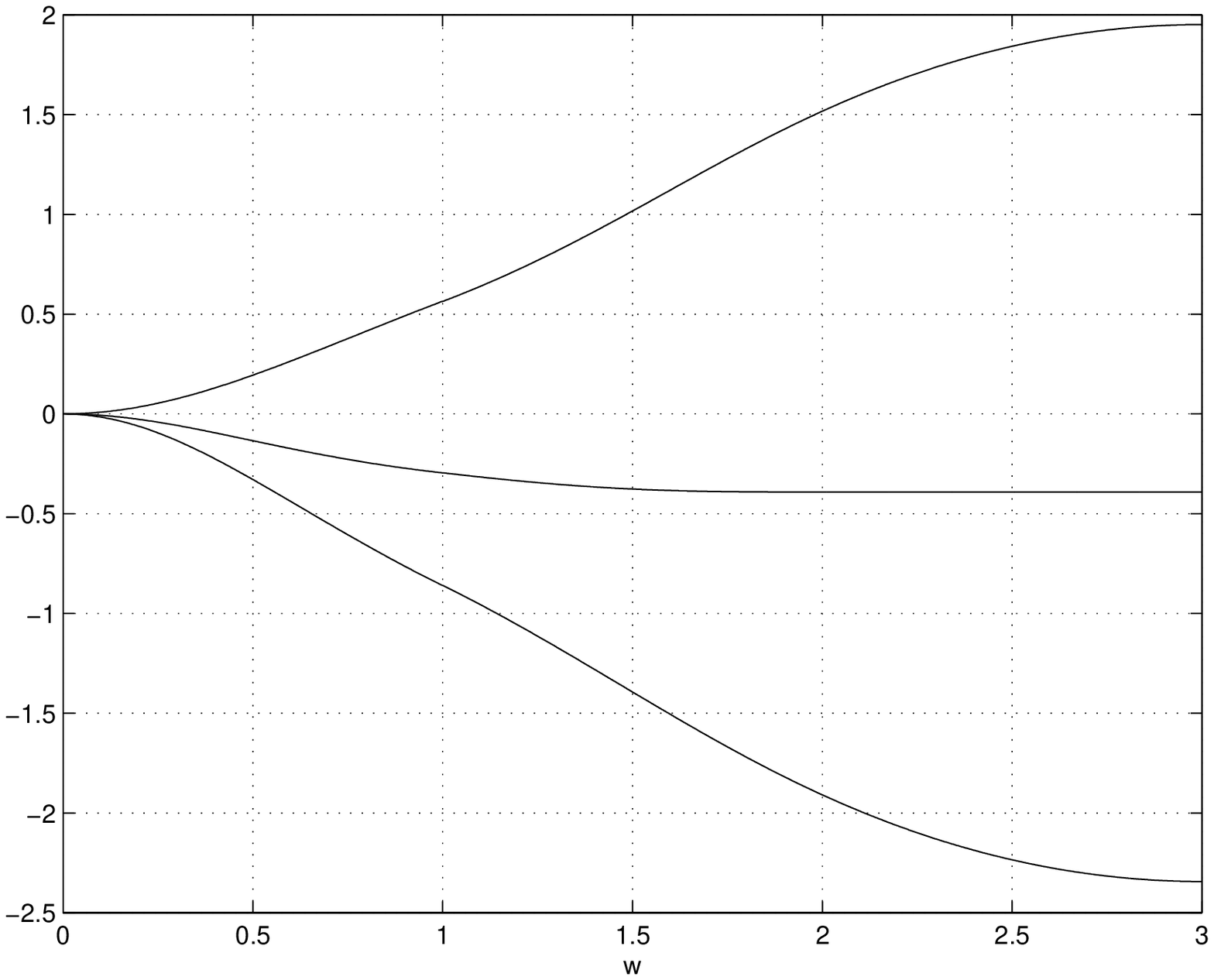}}}
      {\scalebox{0.45}{\includegraphics {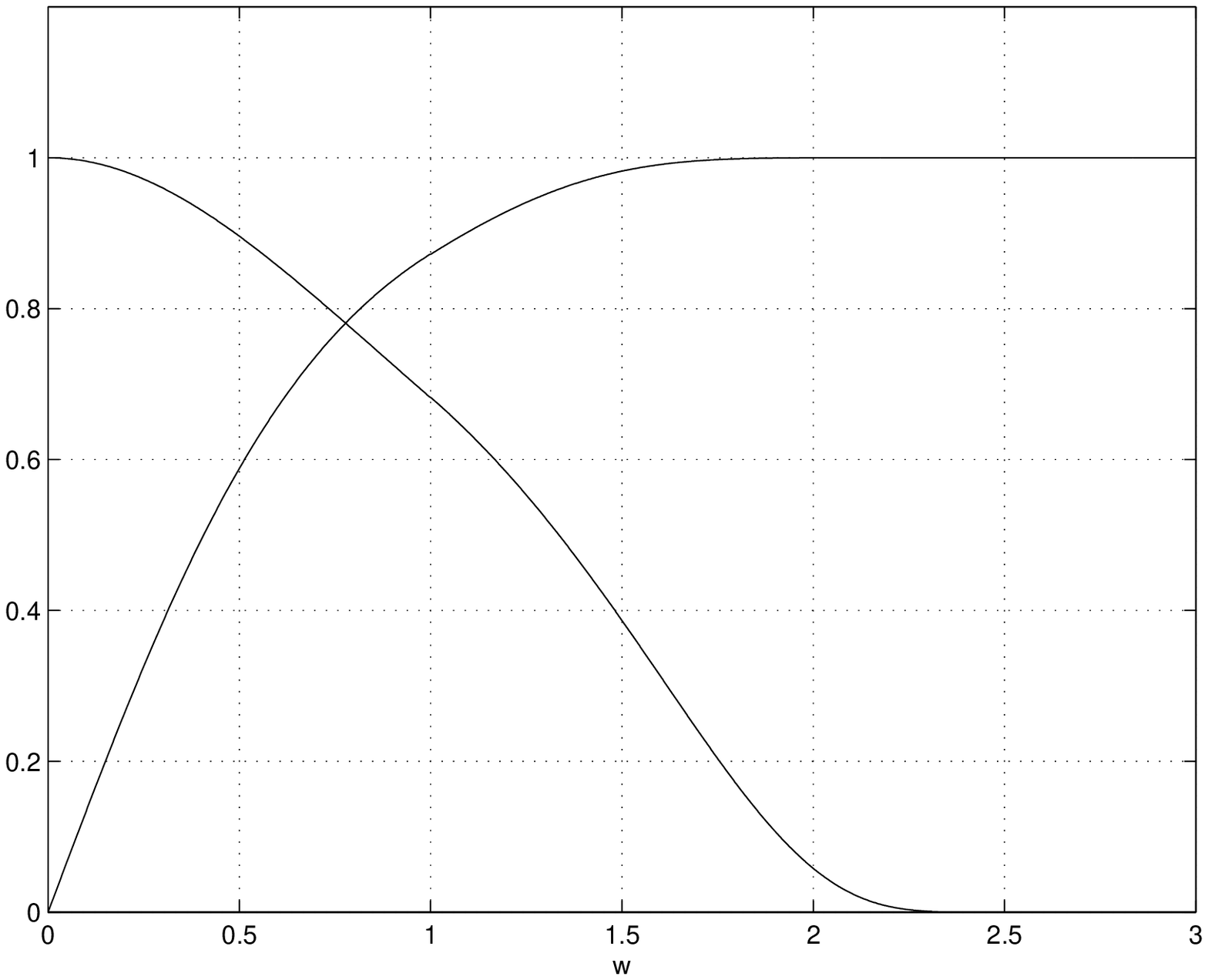}}}}
\put(-325,150){\fontsize{8}{8pt}\selectfont\makebox(0,0)[t]{$\gamma$}}
\put(-325,99){\fontsize{8}{8pt}\selectfont\makebox(0,0)[t]{$\nu-1$}}
\put(-325,62){\fontsize{8}{8pt}\selectfont\makebox(0,0)[t]{$\mu$}}
\put(-129,148){\fontsize{8}{8pt}\selectfont\makebox(0,0)[t]{$X$}}
\put(-129,80){\fontsize{8}{8pt}\selectfont\makebox(0,0)[t]{$P$}}
\caption{Cosmic string coupled to gravity. $\alpha=1$ and $\eta=10^{-3}$. 
On the left the metric variables: $\gamma$, $\nu-1$ and $\mu$ are plotted 
multiplied by $10^5$. On the right $X$ and $P$ are plotted. }
\end{figure}
 
The metric and matter variables for $\alpha=1$ and $\eta=10^{-3}$ are shown in
Fig. 7. We find that the behaviour of $X$ and $P$ is very close  to
that obtained for a static cosmic string in Minkowski spacetime shown
in Fig. 6. For physically realistic values of $\eta$ the metric
variables at infinity are close to their Minkowskian values although
the non-zero value of $\gamma_0=\lim_{r \to \infty}\gamma(r)$
indicates that the spacetime is asymptotically conical, that is
Minkowski spacetime minus a wedge with deficit angle
$\Delta\phi=2\pi(1-e^{-\gamma_0})$. Thus a string with $\alpha=1$ and
$\eta=10^{-3}$ has an angular deficit of about $2\times 10^{-5}$ which
corresponds to a grand unified symmetry breaking scale of about
$10^{16}$ GeV \cite{shellard}. For larger values of $\eta$, however,
the deviation from the Minkowskian case becomes substantial. Close to
critical coupling the deficit angle exceeds $2\pi$ for values of
$\eta$ greater than about $0.28$ which explains why the code converges
well for $\eta \lesssim 0.28$. 

\section{The dynamic Cosmic String coupled to gravity}

In this section we discuss the interaction between the cosmic
string and the gravitational field. Here we will simply outline 
the numerical scheme, the full details are given in paper II. 
The field equations in the $(t,\rho)$ coordinates
are given by equations (\ref{E1})--(\ref{E1a}) 
while those in $(u,y)$ coordinates
are given by equations (\ref{compact1})--(\ref{compact7}). 
In both cases there are two
additional Einstein equations which are consequences of the others and
are used to check the code. In fact two numerical schemes were
employed. The first was an explicit CCM scheme similar to that
employed by Dubal et al. \cite{CDD} and d'Inverno et al. \cite{DDS}. 
However the use of the
geometrical variables $\nu$ and $\tau$ in both the interior Cauchy
region and the exterior characteristic region significantly improves
the interface and results in a genuinely second order scheme with
good accuracy even with both polarizations present. For the vacuum equations 
the code also exhibits long term stability. However when the matter
variables are included this code performs less satisfactorily. This
is because of the existence of exponentially growing non-physical
solutions. It is possible to control these diverging solutions by
multiplying the $u$-derivatives of the matter variables by a
smooth `bump function' which vanishes at $y=0$ but is equal to 1 for
$y>c$ (where $c$ is a parameter). This produces satisfactory results
but the bump function introduces some noise into the scheme
which eventually gives rise to instabilities. 

A  much better
solution is to control the asymptotic behavior at infinity by using
an implicit scheme. The main problem with the system of differential 
equations is the irregularity of the equations at both the origin and
null infinity. Therefore just as in the static case considered in the
previous section  the scheme employed divides the spacetime
into two regions, an inner region $r\leq 1$ where coordinates $(u,r)$
are used and an exterior region $r\geq 1$ where the $(u,y)$ coordinates
are used. The equations in both regions are written as a first order
system connected by an interface and the evolution is accomplished
using a code based on the implicit Crank--Nicholson scheme. 
This implicit scheme provided a simple way of implementing the
boundary conditions and thus circumventing all problems with the 
irregularities.
The outer boundary conditions as well as the equations in the outer region,
the first order system used for the numerics and the interface are given in
paper II. The implicit code showed very good agreement with both the exact
(vacuum) and previously obtained (static) numerical solutions. It also
showed clear second order convergence and very long term
stability. For convenience characteristic coordinates were also used 
in the inner region but we do not think that their use was responsible for
the good features of this code and believe that an implicit CCM scheme
would have produced similar accuracy, convergence and long term stability.

The code is able to consider the interaction of the cosmic string with
a gravitational field with two degrees of freedom, however here we
simply describe the interaction with a Weber--Wheeler wave which has
just one degree of freedom. We consider a pulse which comes in from past null
infinity and interacts with a cosmic string in its static equilibrium configuration. 
This interaction causes the string to oscillate which in
turn affects the gravitational field as measured by $\nu$ and
$\tau$. The oscillations in both $X$ and $P$ decay as one approaches
null infinity (i.e. as $r \to \infty$ for fixed $u$) and
also for fixed $r$ as $u \to \infty$. After the oscillation has died
away the string variables
$X$ and $P$ return to their static values. Note
however that this decay is rather slow and being able to show this
effect depends upon the long term stability of the code. 
Although oscillations are
observed in both $X$ and $P$ the character and frequency of these
oscillations is rather different. 

\begin{figure}[ht] \centering
\mbox{{\scalebox{0.40}{\includegraphics {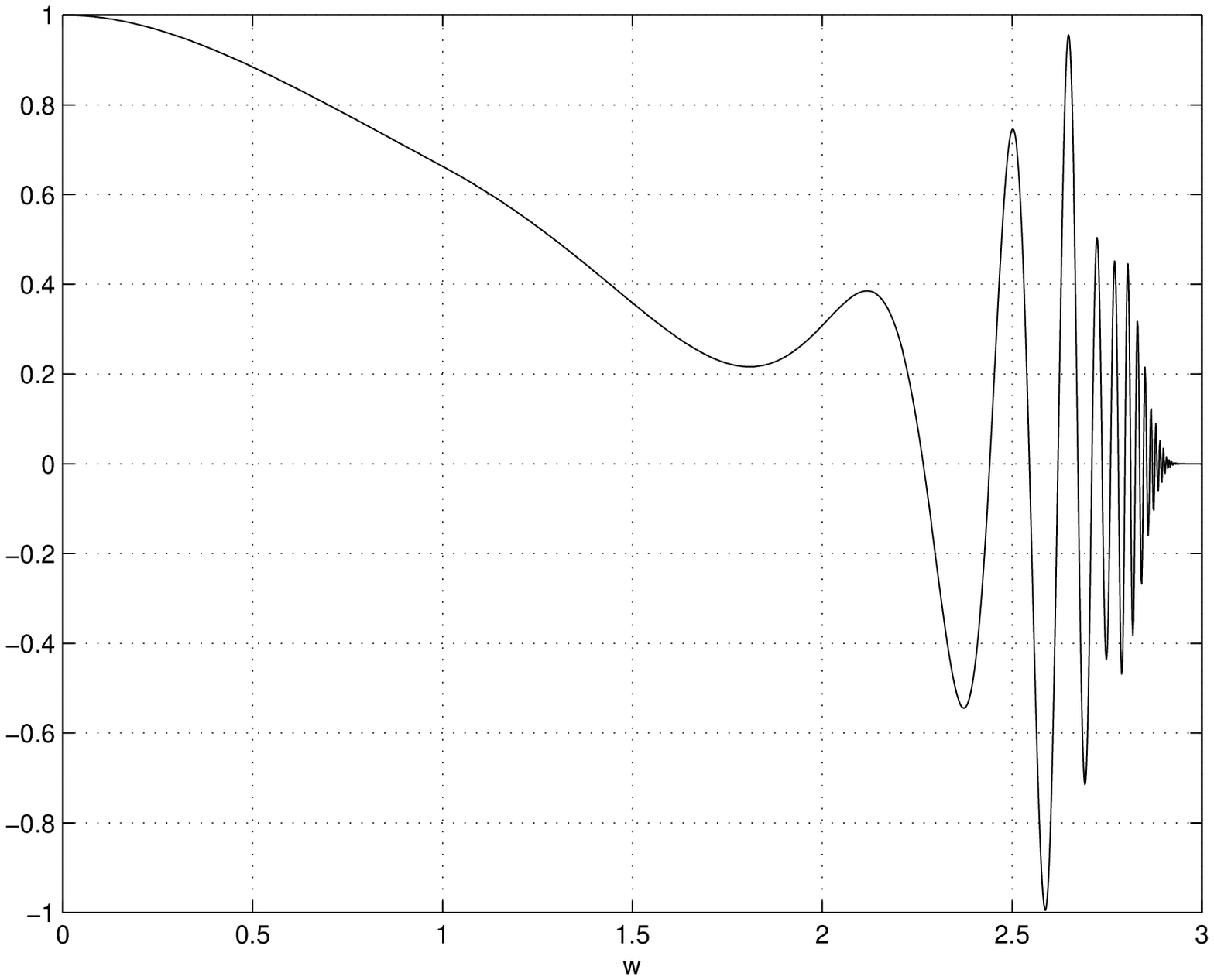}}}
      {\scalebox{0.40}{\includegraphics {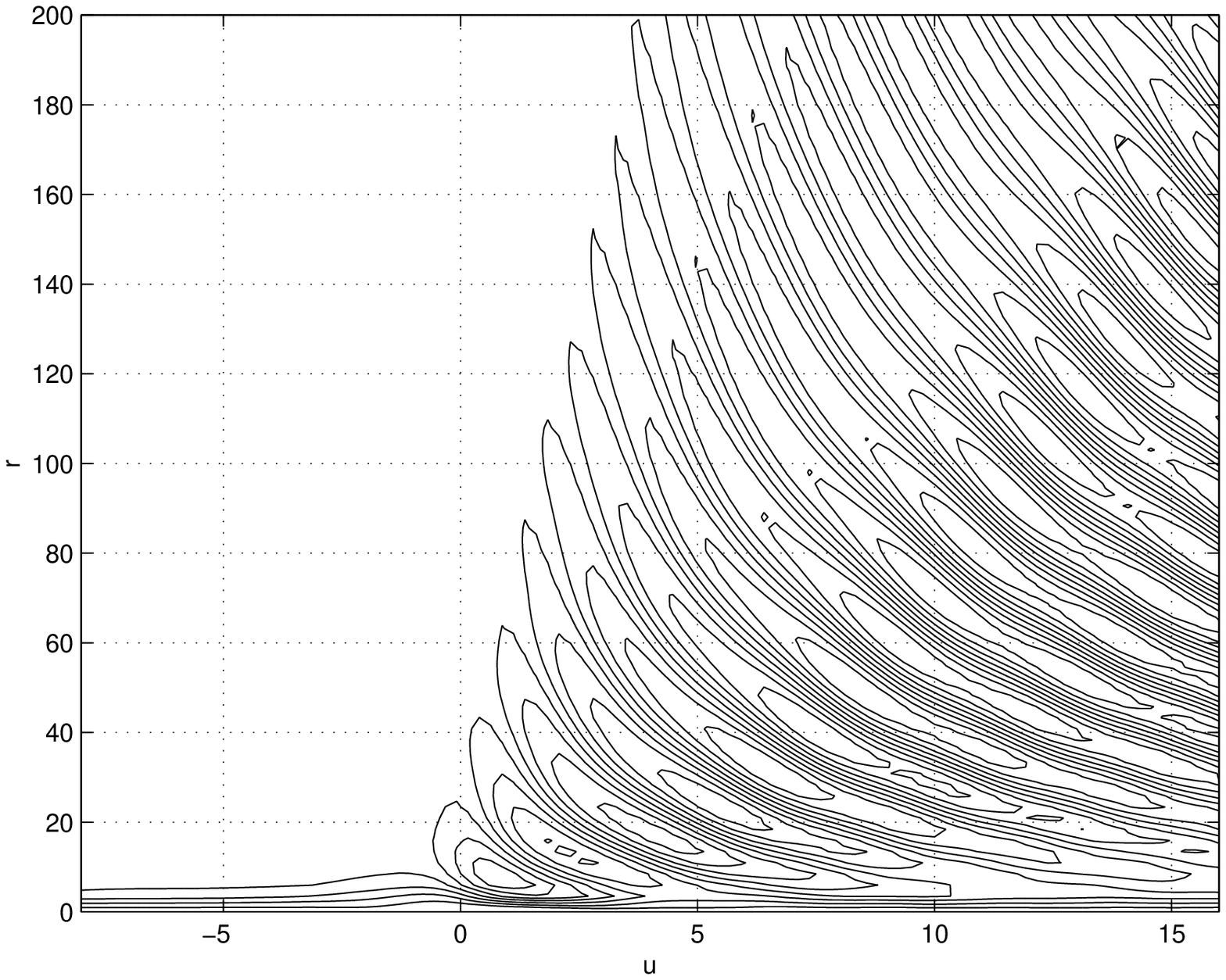}}}}
\caption{On the left: $P(u,w)$ for $\alpha=1$, $\eta=10^{-3}$ and
$u=8.5$. On the right: The corresponding contour plot of $P(u,r)$ for
$0 \leq r \leq 200$ and $-8 \leq u \leq 16$. Note the use of $r$ for values 
greater than one in this plot.}
\end{figure}
\begin{figure}[ht]\centering
\scalebox{0.45}[0.5]{\includegraphics {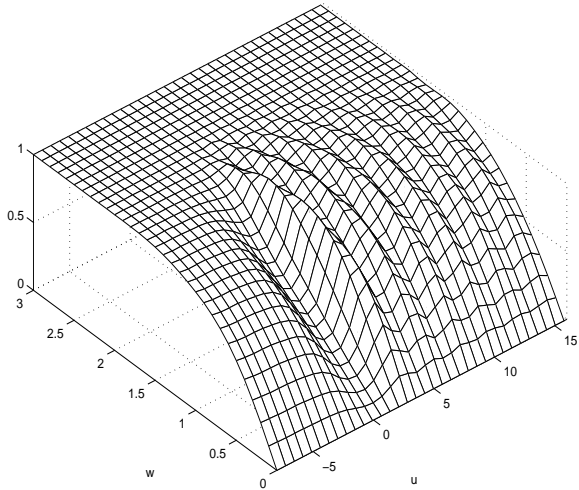}}
\caption{$X(u,w)$ for $\alpha=1$, $\eta=10^{-3}$.}
\end{figure}

In Fig. 8. we plot P for $\alpha=1$ and $\eta=10^{-3}$ at a time
$u=8.5$ (left panel). The oscillations out to large radii can be
clearly seen
. The contour plot on the right shows the ringing behaviour of the
string and the slow decay of the oscillations in $P$. In contrast the
oscillations in $X$ displayed in Fig. 9. are restricted to small radii
and decay on a shorter timescale.

An investigation of the frequencies ${\tilde f}_X$ and ${\tilde f}_P$ of the
oscillations of the scalar field $X$ and the vector field $P$ (in the
rescaled unphysical variables) indicates
that they are relatively insensitive to the value of $\eta$ and the
Weber--Wheeler pulse which excites the string. They are also largely
independent of the radius at which they are measured. However, although
${\tilde f}_X$ is also independent of $\alpha$, we find that ${\tilde
f}_P$ is proportional to $\sqrt{\alpha}$. When we convert to the
physical fields $S$ and $P$ and use the physical coordinates $(t,\rho)$
we find that the frequencies in natural units are given by
\begin{eqnarray}
  f_S &\sim& {\sqrt\lambda\eta} \\
  f_P &\sim& {e \eta}. 
\end{eqnarray}
If we now use the fact that the masses of the scalar and vector fields
are given by $m_S^2 \sim \lambda\eta^2$ and $m_P^2 \sim e^2\eta^2$
\cite{shellard}  this gives
\begin{eqnarray}
  f_S &\sim& {m_S} \\
  f_P &\sim& {m_P}. 
\end{eqnarray}
In fact these frequencies are also obtained by considering the simpler
model of a dynamic cosmic string in Minkowski space with equations of
motion (76)--(77) provided one gives $S$ and $P$ similar initial
conditions to that produced by the interaction with the pulse of
gravitational radiation. Thus the main role of the gravitational field
as far as the string is concerned is to provide a mechanism for
perturbing the string. 
This is discussed more fully in paper II.

\section{Conclusion}

In this paper we have shown how the method of Geroch decomposition may
be used to recast the field equations for a time dependent cylindrical cosmic
string in four dimensions in terms of fields on a reduced 3-dimensional
spacetime. This has the advantage that it has a well defined notion of
null infinity. It is therefore possible to conformally compactify the
3-dimensional spacetime and avoid the need for artificial outgoing
radiation conditions. An additional feature of this approach is that
it naturally introduces two geometrically defined variables $\nu$ and
$\tau$ which encode the two gravitational degrees of freedom in the
original spacetime. 

We have described how the field equations for a cosmic string coupled
to gravity may be solved numerically by dividing the 3-dimensional
spacetime into two regions; an interior region $r \leq 1$, and an
exterior region $r \geq 1$ in which the coordinate $y$ is used. The
use of the geometric variables $\nu$ and $\tau$ greatly simplifies the
transmission of information at the interface $r=1$. Although an
explicit CCM code worked very effectively in the vacuum case, the
asymptotic behavior of the matter fields $S$ and $P$ made it less
effective when the string is coupled to the gravitational field. An
alternative implicit fully characteristic scheme however exhibited good
accuracy and long term stability.

In this paper we have demonstrated the effectiveness of the CCM codes in
reproducing the results of the Weber--Wheeler solution and two vacuum
spacetimes with two degrees of freedom due to Xanthopoulos and Piran
et al. This involved calculating the Geroch potential for these
solutions and using it to compare with the numerically computed
values. We have also described a code for a static cosmic string which uses
a relaxation scheme and provides initial data for the dynamic code.

In the final section of the paper we briefly described the interaction
of the cosmic string with a Weber--Wheeler gravitational wave. The
pulse of gravitational radiation excites the string and causes the
fields $S$ and $P$ to oscillate with frequencies  proportional
to their respective masses. The full details of the code, the
convergence testing and the interaction between the string and the
gravitational field are described in paper II.

\acknowledgements

We would like to thank Ray d'Inverno for helpful discussions and
Denis Pollney for help with GRTensor II.


\end{document}